\documentclass[12pt]{article}
\PassOptionsToPackage{numbers, compress}{natbib}
\usepackage{booktabs,amsfonts,nicefrac,graphicx,enumitem,rotating,mathtools,wrapfig,url,lineno,microtype,subcaption,blindtext,cite,amsmath,amsthm,ulem,bm}
\usepackage[preprint]{neurips_2020}
\usepackage[dvipsnames]{xcolor}

\usepackage[onehalfspacing]{setspace}
\usepackage[ruled,vlined]{algorithm2e}
\usepackage[final]{hyperref}
\usepackage[font=small,labelfont=bf,labelsep=space]{caption}
\hypersetup{
	colorlinks=true,       
	linkcolor=blue,        
	citecolor=blue,        
	filecolor=magenta,     
	urlcolor=blue         
}
\newcommand{\ind}{\perp\!\!\!\!\perp}
\theoremstyle{definition}
\newtheorem{exmp}{Example}[section]

\newtheorem{theorem}{Theorem}

\makeatletter
 \def\@textbottom{\vskip \z@ \@plus 1pt}
 \let\@texttop\relax
\makeatother
\newtheorem{definition}{Definition}
\makeatletter
\renewcommand{\fnum@figure}{Fig \thefigure.}
\makeatother
\title{Statistical Perspective on Functional and Causal Neural Connectomics: A Comparative Study}
\author{
  Rahul Biswas\\ 
  Department of Statistics\\
  University of Washington\\
  Seattle, WA, 98195\\
  \texttt{rbiswas1@uw.edu}\\
  \And
  Eli Shlizerman \\ 
  Department of Applied Mathematics\\
  Department of Electrical \& Computer Engineering\\  University of Washington\\
  Seattle, WA, 98195 \\
  \texttt{shlizee@uw.edu} \\
}

\begin{document}

\maketitle

\begin{abstract}
Representation of brain network interactions is fundamental to the translation of neural structure to brain function. As such, methodologies for mapping neural interactions into structural models, i.e., inference of functional connectome from neural recordings, are key for the study of brain networks.
While multiple approaches have been proposed for functional connectomics based on statistical associations between neural activity, association does not necessarily incorporate causation. Additional approaches have been proposed to incorporate aspects of causality to turn functional connectomes into causal functional connectomes, however, these methodologies typically focus on specific aspects of causality. This warrants a systematic statistical framework for causal functional connectomics that defines the foundations of common aspects of causality. Such a framework can assist in contrasting existing approaches and to guide development of further causal methodologies. In this work, we develop such a statistical guide. In particular, we consolidate the notions of associations and representations of neural interaction, i.e., types of neural connectomics, and then describe causal modeling in the statistics literature. We particularly focus on the introduction of directed Markov graphical models as a framework through which we define the Directed Markov Property -- an essential criterion for examining the causality of proposed functional connectomes. We demonstrate how based on these notions, a comparative study of several existing approaches for finding causal functional connectivity from neural activity can be conducted. We proceed by providing an outlook ahead regarding the additional properties that future approaches could include to thoroughly address causality.
\end{abstract}

\section{Introduction}
The term ``connectome" typically refers to a network of neurons and their anatomical links, such as chemical and electrical synapses. The connectome represents the anatomical map of the neural circuitry of the brain \citep{sporns2005human}. Connectome mapping can be achieved with the help of imaging techniques and computer vision methods at different scales~\citep{shi2017connectome, sarwar2020towards,xu2020connectome}. The aim of finding the connectome is to provide insight into how neurons are connected and how they interact to form brain function. 

While the anatomical connectome includes the backbone information on possible ways of how neurons could interact, it does not fully reflect the ``wiring diagram of the brain", which is expected to incorporate the dynamic nature of neurons' activity and their interactions~\citep{lee2011specificity,kopell2014beyond,kim2017neural,kim2018movingcelegans}. In particular, the anatomical connectome does not correspond to how the anatomical structure relates to brain function since each anatomical connectivity map can encode several functional outcomes of the brain \citep{bargmann2013connectome}. Thereby, the term connectome has been extended beyond the anatomical meaning. In particular, a map that reflects neurons functions is named Functional Connectome(FC) and it represents the network of associations between neurons with respect to their activity over time \citep{reid2012functional}. Finding FC is expected to lead to more fundamental understanding of brain function and dysfunction \citep{hassabis2017neuroscience}. Indeed, FC is expected to include and facilitate inference of the governing neuronal pathways essential for brain functioning and behavior~\citep{finn2015functional}. Two neurons are said to be functionally connected if there is a significant relationship between their activity over time where the activity can be recorded from neurons over time and measured with various measures~\citep{shlizerman2012neural}. In contrast to the anatomical connectome, the functional connectome needs to be inferred, as it cannot be directly observed or mapped, since the transformation from activity to associations is intricate.

Several approaches have been introduced to infer the FC. These include approaches based on measuring correlations, such as pairwise correlation~\citep{rogers2007assessing,preti2017dynamic}, or sparse covariance matrix that is comparatively better than correlations given limited time points~\citep{xu2015dynamic,wee2016diagnosis}.
Furthermore, for such scenarios, regularized precision matrix approaches were proposed to better incorporate conditional dependencies between neural time courses, where the precision matrix is inferred by a penalized maximum likelihood to promote sparsity \citep{varoquaux2010brain,smith2011network,friedman2008sparse}. While there is a wide variety of methods, there is still a lack of unification as to what defines the ``functional connectome". A taxonomy which provides a systematic treatment grounded from definitions and followed with algorithmic properties is currently unavailable.

Moreover, the prevalent research on FC, outlined above, deals with finding associations between neural signals in a non-causal manner. That is, in such a mapping we would know that a neuron A and a neuron B are active in a correlated manner, however, we would not know whether the activity in neuron $A$ causes neuron $B$ to be active ($A\rightarrow B$), or is it the other way around ($B\rightarrow A$)? Or, is there a neuron $C$ which intermediates the correlation between $A$ and $B$ ($A\leftarrow C\rightarrow B$)? In short, questions of this nature distinguish causation from correlation.

In this work, we provide a statistical framework for FC and investigate the aspect of causality in the notion of Causal Functional Connectome (CFC) which would answer the aforementioned causal questions \citep{valdes2011effective,ramsey2010six}. Specifically, we introduce the directed Markov graphical models as a framework for the representation of functional connectome and define the Directed Markov Property -- an essential criterion for examining the causality of proposed functional connectomes. The framework that we introduce allows us to delineate the following properties for a statistical description of causal modeling
\begin{enumerate}
\item Format of causality.
\item Inclusion of temporal relationships in the model.
\item Generalization of the statistical model. 
\item Dependence on parametric equations.
\item Estimation-based vs. hypothesis test-based inference of CFC from recorded data.
\item Inclusion of cycles and self-loops.
\item Incorporation of intervention queries in a counterfactual manner.
\item Ability to recover relationships between neurons when ground-truth dynamic equation of neural activity are given. 
\end{enumerate}
We discuss the applicability and the challenges of existing approaches for CFC inference with respect to these statistical properties. In particular, we compare existing approaches for causal functional connectome inference, such as Granger Causality (GC), Dynamic Causal Modeling (DCM) and Directed Probabilistic Graphical Models (DPGM) based on these properties. The comparative study provides a taxonomy of causal properties that existing methods address and an outlook of the properties for future extensions to address.

\section{Neural Connectomics: Anatomical and Functional}
Recent advances in neuro-imaging has made it possible to examine brain connectivity at micro and macro scales. These, for example, include electron microscopy reconstruction of the full nervous system with neurons and synapses of \textit{C. elegans} in the mid-1980s \citep{white1986structure}. Recent non-invasive diffusion tensor imaging, followed by computational tractography, allow to recover fiber tract pathways in human brain \citep{conturo1999tracking,le2003looking,catani2003occipito}. Also, two-photon tomography facilitates imaging axonal projections in the brain of mice \citep{ragan2012serial}. Although the anatomical reconstruction gives insights into the building blocks and wiring of the brain, how that leads to function remains unresolved. A representative example is \textit{C. elegans}, in which connections and neurons were fairly rapidly mapped and some neurons were associated with functions, it is still unclear what most of the connections do \citep{morone2019symmetry}. 

It became increasingly clear that anatomical wiring diagram of the brain could generate hypotheses for testing, but it is far from the resolution of how the anatomical structure relates to function and behavior. This is because each wiring diagram can encode several functional and behavioral outcomes \citep{bargmann2013connectome}. In both vertebrate and invertebrate brains, pairs of neurons can influence each other through several parallel pathways consisting of chemical and electrical synapses, where the different pathways can either result in similar or dissimilar behavior. Furthermore, neuromodulators re-configure fast chemical and electrical synapses and have been shown to act as key mediators for various brain states. Given this background, it is generally agreed upon that the anatomical synaptic connectivity does not provide adequate information to predict the physiological output of neural circuits. To understand the latter, one needs to understand the flow of information in the network, and for that, there is no substitute for recording neuronal activity and inferring functional associations from it. This is what the functional connectome aims to achieve.

The functional connectome is the network of information flow between neurons based on their activity and incorporates the dynamic nature of neuronal activity and interactions between them. To obtain neuronal activity and dynamics, the neuronal circuit needs to be monitored and/or manipulated. Recent approaches to record such activity include brain wide two-photon calcium single neuron imaging in vivo of \textit{C. elegans} \citep{kato2015global}, wide-field calcium imaging of regions of the brain of behaving mice \citep{zatka2020perceptual}, two-photon calcium imaging \citep{villette2019ultrafast}, Neuropixel recordings of single neurons in the brain of behaving mice \citep{steinmetz2019distributed, steinmetz2018challenges}, and functional Magnetic Resonance Imaging (fMRI) recordings of voxels in the human brain as part of the Human Connectome Project \citep{stocco2019analysis, van2012human}. 

In this section, we set the notions of the neural connectome with regards to anatomical and functional aspects. These notions are from a statistical perspective and will assist us to connect connectomics in further sections with causation.


\subsection{Anatomical Connectome}
Let us consider a brain network $V=\{v_1,\ldots, v_N\}$ with $N$ neurons labeled as $v_1,\ldots,v_N$. We will denote the edges as $E_a \subset V\times V$ between pairs of neurons that correspond to anatomical connectivity between the neurons \citep{sporns2005human}. We will refer to the graph $G=(V,E_a)$ as the anatomical connectome between the neurons in $V$. Each edge $(v,u) \in E_a$ will be marked by a weight $w_{vu}\in \mathbb{R}$ that quantifies the strength of the anatomical connection from $v$ to $u$.
\subsubsection*{Examples}
\paragraph{2.1.1) Binary Gap Connectome.} If there is a gap junction connection from neuron $v$ to neuron $u$, the set $E_a$ will include the edges $v\rightarrow u$ and $u\rightarrow v$ and they will be marked with weight $1$ \citep{jarrell2012connectome}. The resulting graph is undirected in the sense that $(v,u)\in E_a$ iff $(u,v) \in E_a$. The resulting weight matrix with entries $w_{vu}$ is symmetric.
\paragraph{2.1.2) Weighted Synaptic Connectome.}If there is a synaptic connection from neuron $v$ to neuron $u$, the set $E_a$ will include the edge $v\rightarrow u$ and it will be marked with a weight which is equal to the number of synaptic connections starting from neuron $v$ to neuron $u$. The resulting graph is a directed graph in the sense that $(v,u) \in E_a$ does not imply $(v,u)\in E_a$, and the weight matrix $(w_{vu})$ is asymmetric \citep{varshney2011structural}.




\subsection{Functional Connectome (Associative)}\label{sec:assoc-fc}
\emph{Functional Connectome} (FC) is a graph with nodes being the neurons in $V$ and pairwise edges representing a stochastic relationship between the activity of the neurons. Weights of the edges describe the strength of the relationship. Let $X_v (t) \in \mathbb{R}$ denote a random variable measuring the activity of neuron $v$ at time $t$. Examples for such variables are instantaneous membrane potential, instantaneous firing rate, etc. 
\emph{Associative FC} (AFC) is an undirected graph, $G=(V,E)$, where edges, $E_{v,u} \in E$, an undirected edge between $v$ and $u$, represents stochastic association between neuron $v$ and $u$. 
Edge weights signify the strength of the associations. Different approaches define stochastic associations leading to different candidates for AFC, as follows.

\subsubsection{Pairwise Associative Connectivity}
We first describe pairwise stochastic associations (\emph{pairwise AFC}).
Let us consider recordings at time points $0,\ldots, T$, with activities $X_v=\{X_{v}(t): t\in 0,\ldots,T\}$ for neuron $v$. The following measures of pairwise association will correspond to pairwise AFC.

\paragraph{Pearson's Correlation} The Pearson's correlation between $X_v$ and $X_u$ is defined as \[r(X_v,X_u)=\frac{\sum_{t=0}^T (X_v (t) - \bar{X}_v)(X_u (t) - \bar{X}_u)}{\sqrt{\sum_{t=0}^T (X_v (t) - \bar{X}_v)^2 \sum_{t=0}^T(X_u (t) - \bar{X}_u)^2}}\] where $\bar{X}_v = \frac{1}{T+1} \sum_{t=0}^T X_v(t)$, for $v\in V$. Pearson's correlation takes a value between -1 and 1 and the further its value is from 0, the larger is the degree of pairwise association. Neurons $v$ and $u$ are connected by pairwise AFC with respect to Pearson's correlation if $r(X_v,X_u)$ is greater than a threshold in absolute value and the value $r(X_v,X_u)$ would be the weight of the connection, i.e. $E_{v,u} = \text{thresh} \left(r(X_v,X_u)\right)$. 
Correlation coefficient and the AFC based on it are sensitive to indirect third party effects such as an intermediary neuron, poly-synaptic influences, indirect influences, and noise.

\paragraph{Partial Correlation} Partial correlation is an additional measure of pairwise stochastic association between random variables defined as follows. Let the covariance between $X_v$ and $X_u$ be $C_{vu}=\frac{1}{T+1}\sum_{t=0}^T (X_v (t) - \bar{X}_v)(X_u (t) - \bar{X}_u)$. The matrix of covariances, $\Sigma = (C_{vu})_{1\leq v,u \leq N}$ is called the covariance matrix for the variables $X_1,\ldots, X_N$. Let the $v,u$-th entry of the inverse covariance matrix $\Sigma^{-1}$ be $\gamma_{vu}$. The partial correlation between $X_v$ and $X_u$ is defined as \[\rho(X_v,X_u)=-\frac{\gamma_{vu}}{\sqrt{\gamma_{vv}\gamma_{uu}}}\]Partial correlation rectifies the problem of correlation coefficient being sensitive to third party effects since it estimates the correlation between two nodes after removing the variance shared with other system elements. Partial correlation takes a value between -1 and 1 and the further its value is from 0, the larger is the degree of pairwise association. Neurons $v$ and $u$ are connected by pairwise AFC with respect to partial correlation, if $\rho(X_v, X_u)$ is greater than a threshold in absolute value and the value $\rho(X_v,X_u)$ would be the weight of the connection, i.e. $E_{v,u}=\text{thresh}\left(\rho(X_v,X_u)\right)$. 

\subsubsection{Undirected Probabilistic Graphical Model}
\begin{figure}[t!]
    \centering
    \includegraphics[width = \textwidth]{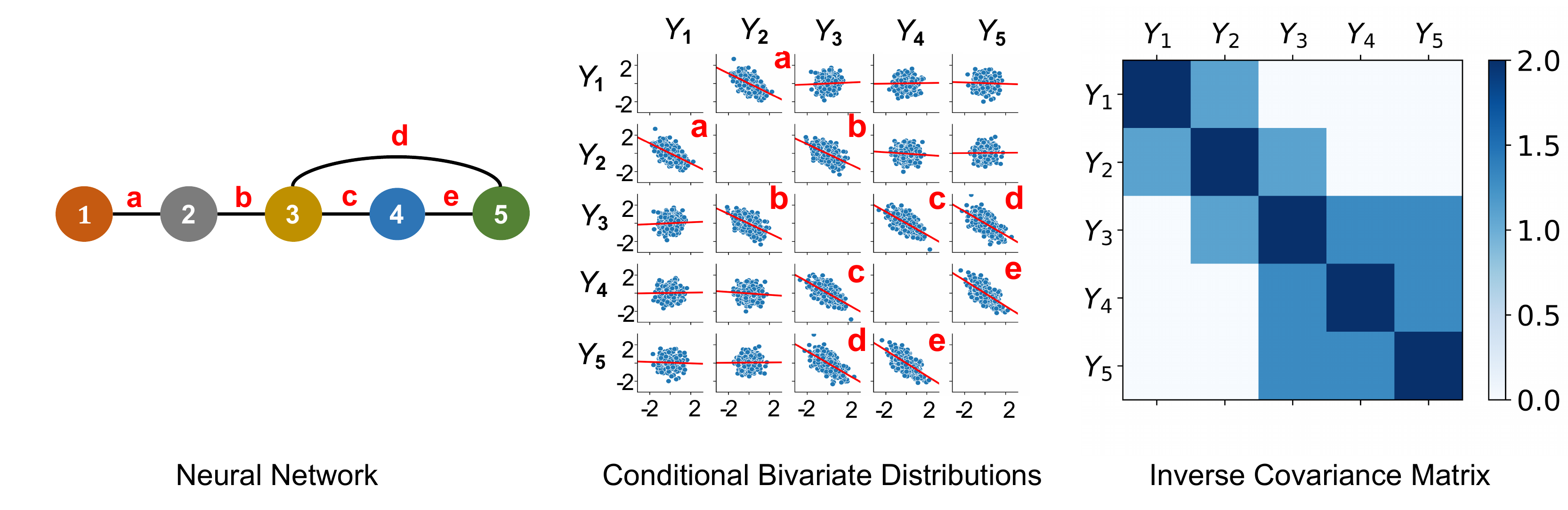}
    \caption{\textbf{An undirected PGM and its Markov Property.} Left: Network of 5 neurons defined in Example \ref{ex:undir-gm}. Neurons are labeled as $1-5$ and edges are labeled as a-e. Middle: $Y_1,\ldots,Y_5$ following centered multivariate Gaussian distribution with entries of inverse covariance matrix $\gamma_{ij}$ such that $\gamma_{13}=\gamma_{14}=\gamma_{15}=\gamma_{24}=\gamma_{25}=0$, when it factorizes with respect to the Neural Network. This plot shows conditional bivariate distributions given other variables $\in (-0.2,0.2)$ and demonstrates Eq. (\ref{eq:ugmarkov}) where $Y_v$ and $Y_u$ are not correlated conditional on other variables (seen by nearly flat red trend-lines) for $(v,u)$ not an edge of the Neural Network. This indicates that $Y_1,\ldots,Y_5$ satisfies the Markov Property with the Neural Network. Right: Due to a Gaussian distribution, the non-zero entries of the Inverse Covariance Matrix of $Y_1,\ldots, Y_5$ correspond to the edges of the Neural Network.}
    \label{fig:undirgm}
\end{figure}

Undirected Probabilistic Graphical Models (PGM) allow for modeling and infering stochastic associations while considering multi-nodal interactions beyond pairwise manner through a graphical model. Let $G=(V,E)$ be an undirected graph  with neuron labels $V=\{v_1,\ldots,v_N\}$ and edges $E$ (Figure \ref{fig:undirgm}). Let $Y_v$ denote a scalar-valued random variable corresponding to $v\in V$, for example, $Y_v$ can be the value of $X_v$ at recording time $t$: $Y_v=X_v(t)$, or average of recordings over time: $Y_v=\bar{X}_v$, etc. For a set of neurons $A\subset V$, let $\bm{Y}_A$ denote the random vector $(Y_v,v\in A)$. The random vector $\bm{Y}_V$, is said to satisfy the \emph{undirected graphical model} with graph $G$, if, $Y_v$ is conditionally independent of $Y_u$ given $\bm{Y}_{V\setminus \{v,u\}}$ for $(v,u) \not\in E$, denoted as
\begin{equation}\label{eq:ugmarkov}Y_v \ind Y_u \vert \bm{Y}_{V\setminus \{v,u\}} \text{ for } (v,u) \notin E.\end{equation}
When Eq. (\ref{eq:ugmarkov}) holds, $Y_v$ is said to satisfy the \emph{Markov property} with the undirected graph $G$ \citep{wainwright2008graphical}. The Markov property with undirected graph $G$ translates each \emph{absent} edge between a pair of nodes $v$ and $u$ into \emph{conditional independence} of $Y_v$ and $Y_u$ given all other nodes. In other words, nodes in the undirected PGM that are not connected by an edge appear as non-interacting when all the nodes are examined.

\begin{definition}\label{def:unirectedPGMFC}
The \emph{AFC} for neurons in $V$ is the undirected graph $G=(V,E)$ such that $(Y_v,v\in V)$ satisfies the \emph{Markov Property} with respect to $G$. When $Y_v=X_v(t)$, the associative FC is contemporaneous at each time $t$.
\end{definition}

According to this definition, the graph edges $E_{v,u}$ are well-defined. With respect to their weights, there are no unique candidates. When $(Y_v, v\in V)$ follows a multivariate Gaussian distribution, an edge is present if and only if the partial correlation between them is non-zero. Thereby, in practice, partial correlation $\rho(X_v, X_u)$ is typically used for weight of edge $E_{v,u} \in E$ (Figure \ref{fig:undirgm}).



In the above we have defined AFC to be the undirected graphical model that has the Markov Property. A natural question that arises is what kind of probability distributions for the graphical model allow it to follow the Markov Property? The Factorization Theorem by Hammersley, Clifford and Besag prescribes conditions on the probability distributions of $Y_v, v\in V$ for which undirected graphical model has the Markov property \citep{drton2017structure}.
\begin{theorem}[Factorization Theorem]\label{thm:hcliff}
If $Y_v, v \in V$ has a positive and continuous density $f$ with respect to the Lebesgue measure or is discrete with positive joint probabilities, then it satisfies the Markov property (Eq. \ref{eq:ugmarkov}) with respect to $G=(V,E)$ if and only if the distribution of $Y_v, v\in V$ \emph{factorizes} with respect to $G$, which means,
\begin{equation}\label{eq:facto}
f(y) = \prod_{C\subset G: C \text{ is complete} } \phi_C (y_C), y\in \mathbb{R}^V
\end{equation}
where, $f$ is the density of $Y_v, v\in V$, $\phi_C$ is an arbitrary function with domain $\mathbb{R}^C$, and $y_C$ is the sub-vector $(y_v : v \in C)$ and $C\subset G$ is complete, i.e. $E_{v,u}\in E$ for all $v\neq u \in C$, are connected by an edge.
\end{theorem}

Under the multivariate Gaussian assumption, Theorem \ref{thm:hcliff} yields a simple prescription for obtaining the undirected graph with the Markov Property. When $Y_v, v\in V$ are distributed as multivariate Gaussian with positive definite covariance matrix $\Sigma$, then $G$ is determined by the zeroes of the inverse covariance matrix $\Sigma^{-1}$, i.e., $E_{v,w}\in E$ iff $\Sigma^{-1}_{ij} \neq 0$. This is illustrated in Example \ref{ex:undir-gm} and Figure \ref{fig:undirgm}. This has been used for inferring undirected PGMs in several applications \citep{epskamp2018gaussian, dyrba2020gaussian}. In such a case, estimation of $G$ is tantamount to estimation of $\Sigma^{-1}$. Methods for estimation of $\Sigma^{-1}$ include Maximum Likelihood Estimation (MLE) which provides non-sparse estimates of $\Sigma^{-1}$ \citep{mlegaussiangraphical} and penalized MLE, e.g. by Graphical Lasso, which provides sparse estimates of $\Sigma^{-1}$, with the estimates being statistically consistent under assumptions \citep{shojaie2010penalized,banerjee2008model,rothman2008sparse,meinshausen2006high}.

\begin{exmp}[Markov Property and Multivariate Gaussian Assumption]\label{ex:undir-gm} Let us consider an example with $5$ neurons. Suppose $(X_1 (t),\ldots, X_5 (t))$ follow a centered multivariate Gaussian distribution with positive definite covariance matrix $\Sigma$ and independent copies over time $t$. Let $\Sigma^{-1} = (\gamma_{ij})_{1\leq i,j \leq N}$ be the inverse covariance matrix. The probability density of $(X_1 (t),\ldots, X_5 (t))$ is given by
\begin{equation}\label{eq:ex1}
f(x^t) \propto \exp \left(-\frac{1}{2}\sum_{i,j=1}^5 \gamma_{ij}x_{i}^t x_{j}^t \right), x^t \in \mathbb{R}^5.
\end{equation}

The neural network graph is illustrated in Figure \ref{fig:undirgm}-left and we note that the graph has the following complete subsets \[\{1,2\},\{2,3\},\{3,4\},\{4,5\},\{3,5\},\{3,4,5\}.\]Thereby, for each $t$, it is observed that 
$\gamma_{13} = \gamma_{14}$ $=\gamma_{15}=\gamma_{24}=\gamma_{25} = 0$ iff the density in Eq. (\ref{eq:ex1}) factorizes as 

\begin{align*}f(x^t)\propto& \exp\left(-\frac{1}{2}({x_1^t}^2+x_1^t x_2^t+{x_2^t}^2)\right)\exp\left(-\frac{1}{2}x_2^t x_3^t\right)\exp\left(-\frac{1}{2}({x_3^t}^2+x_3^t x_4^t+{x_4^t}^2)\right)\\
&\hspace{1.5in}\exp\left(-\frac{1}{2}(x_4^t x_5^t+{x_5^t}^2)\right)\exp\left(-\frac{1}{2}x_3^t x_5^t\right)
\end{align*} 

That is, the density in Eq. (\ref{eq:ex1}) factorizes according to Eq. (\ref{eq:facto}) with respect to the graph. Hence, according to Theorem \ref{thm:hcliff} it follows that $(X_1(t),\ldots, X_5(t))$ satisfies the Markov Property with respect to the graph in Figure \ref{fig:undirgm}-left.
\end{exmp}
\section{From Association to Causation}
\begin{figure}[t!]
    \centering
    \includegraphics[width = 0.9\textwidth]{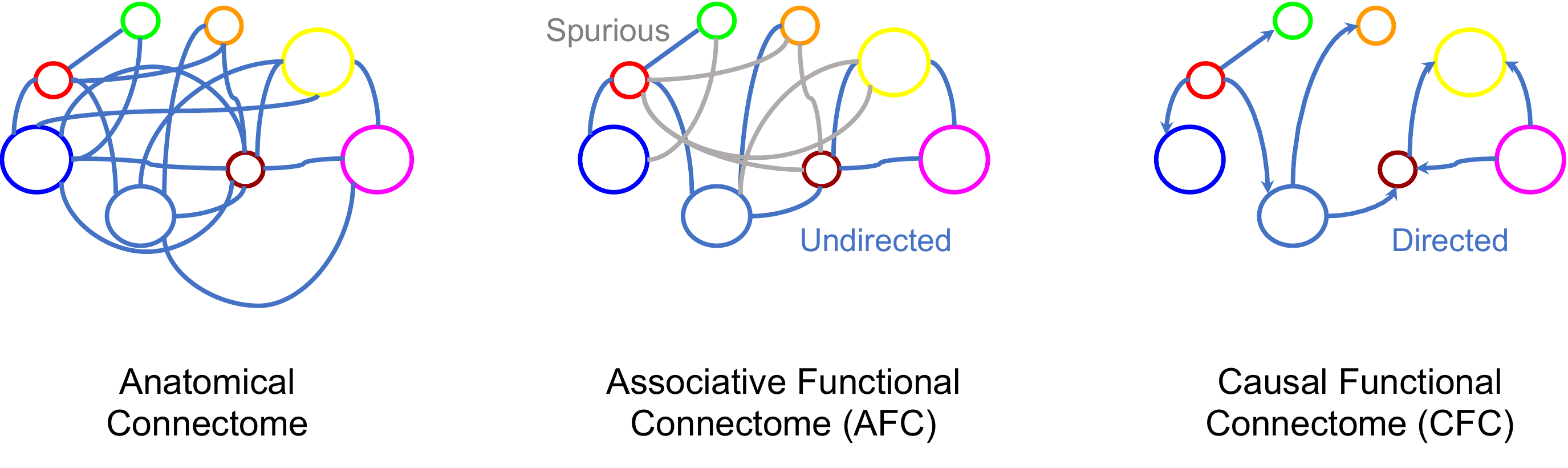}
    \caption{\textbf{Various connectome representations for the same network of neurons.} Left to right: Anatomical connectivity between brain regions, an undirected graph representing AFC where gray indicates spurious edges, and a directed graph with directions representing causal relationships.}
    \label{fig:associative-causal}
\end{figure}

In Section \ref{sec:assoc-fc} we provided a systematic exposition of AFC, however, the ultimate phenomenon of functional connectivity is to capture the causal interaction between neural entities \citep{valdes2011effective,ramsey2010six, friston2011functional, horwitz2003elusive}. Indeed, FC research is already targeting addition of causal statements to associative FC by identifying correlated brain regions as causal entities, suggesting the need to incorporate functional connectivity beyond association \citep{yeo2011organization,power2011functional,power2013control,smith2009correspondence}. From associations alone, causal inference is ambiguous with possible parallel, bidirectional and spurious causal pathways. Narrowing the space of causal pathways inferred from brain signals can significantly progress FC towards its aims of finding causal neural interactions (See Figure \ref{fig:associative-causal}). For example, in a neural circuit, an edge between neuron A and neuron B could mean either neuron A influences neuron B or vice versa. This directionality of influence is unclear from the association. In the instance when neuron C influences activity of both A and B neurons, while they do not influence each other, a spurious association would be found between A and B in the AFC (See example in Figure \ref{fig:associative-causal}). Recent progress in causal inference in statistics literature has made it possible to define causality and make causal inferences. Given this background, we review causal modeling in statistics to aid the development of a novel framework for \textit{Causal Functional Connectivity} (CFC) from neural dynamics.

\subsection{Causal Modeling} 
\begin{figure*}[t!]
    \centering
    \includegraphics[width=0.8\textwidth]{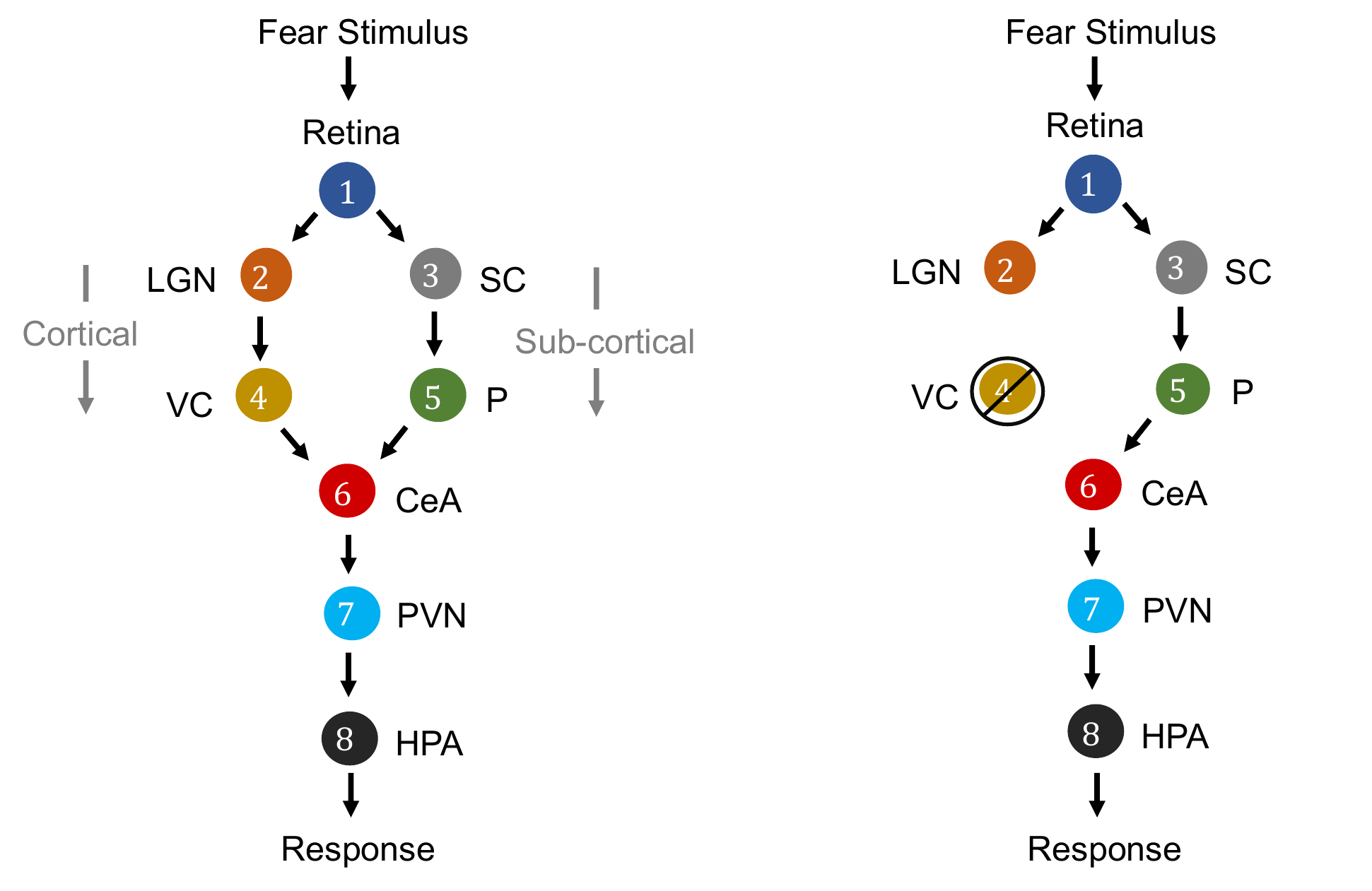}
    \caption{\textbf{Causal graphs for electrical activity of brain nuclei for fear perception from visual stimulus.} Left: Literature describes two routes for fear stimulus propagation from Retina to HPA: Retina $\rightarrow$ LGN $\rightarrow$ VC $\rightarrow$ CeA $\rightarrow$ PVN  $\rightarrow$ HPA (cortical route) and Retina $\rightarrow$SC $\rightarrow$ P $\rightarrow$ CeA $\rightarrow$ PVN $\rightarrow$ HPA (subcortical route). Right: It is demonstrated that even if there is intervention by ablation or lesion in the striate cortex of VC, thereby blindness, yet fear response to visual stimuli ("blindsight") is yielded through the subcortical route \citep{morris1999subcortical,carr2015ll}.}
    \label{fig:causal-example}
\end{figure*}

Causation is a relation between two events, in which one event is the cause and the other is the effect. For example, the sprinkler turned on causes the pavement to get wet. In the context of neuroscience, causality is a major factor. For example, in fear perception in the human brain, the causal relationships among the activity of retina, lateral geniculate nucleus (LGN), superior colliculus (SC), pulvinar (P), central nucleus of the amygdala (CeA), paraventricular nucleus (PVN) and hypothalamus-pituitary-adrenal axis (HPA) can be considered (see Figure \ref{fig:causal-example}). While the information flows from the Retina to HPA, there is no direct link between them. Fear stimulus of the retina causes trigerring of the HPA (response region) mediated by the activity of intermediate brain nuclei in a specific sequence of activation through two merging pathways \citep{carr2015ll,pessoa2010emotion,bertini2013blind}. 

A causal model relates the cause and effect, rather than recording correlation in the data and allows the investigator to answer a variety of queries such as \emph{associational queries} (e.g. having observed activity in LGN, what activity can we expect in CeA?), \emph{abductive queries} (e.g. what are highly plausible explanations for active CeA during fear stimulus?), and \emph{interventional queries} (e.g. what will happen to the causal pathways if there is an ablation of VC?). Often interventional queries are especially of interest in neural connectomics, with interventions including neural ablation (see Figure \ref{fig:causal-example}b) and external neuromodulation \citep{bargmann2013connectome,horn2020opportunities}. In such cases, causal modeling aims to correctly predict the effect of an intervention in a counterfactual manner, that is, without necessarily performing the intervention but from observational data~\citep{pearl2009causality}.

\subsubsection{Representing Causal Relations with a Directed Graph} 
Directed graphs provide convenient means for expressing causal relations among the variables. The vertices correspond to variables, and directed edges between the vertices represent a causal relationship that holds between pairs of variables. Formally, a graph consists of a set $V$ of vertices (nodes) and a set $E \subset V\times V$ of edges that connect some pairs of vertices. Each edge can be either directed or undirected. In a directed graph, all edges are directed. A directed acyclic graph (DAG) is a directed graph without cycles. The directed graph representing causal relationships is called a causal graph. Figure \ref{fig:causal-example} is a causal graph among eight variables representing electrical activity in eight brain regions. If G is a causal graph and there is a directed path from nodes $a$ to $b$ it implies that, the variable corresponding to node $a$ is a cause for the variable corresponding to node $b$. For example, the electrical activity in Retina (node $1$) is a common cause for electrical activity in LGN (node $2$) and SC (node $3$) in Figure \ref{fig:causal-example} (left).

If there is any intervention of one of the variables such as ablation or neuromodulation of nodes, the network topology can be adapted with minor changes. In Figure \ref{fig:causal-example} (right), to represent an ablation of visual cortex (VC), one would delete from the network all links incident to and originating from the node VC. To represent control of activity of VC by neuromodulation, one would delete all edges only incident to VC as then VC is not causally influenced by it's parent region's activity but by external neuromodulation.

\subsubsection{Statistical Properties of CFC Modeling}
In the following we outline several statistical properties that are relevant in the context of causal modeling of functional connectivity.
\begin{enumerate}
\item \emph{Format of causality}. This specifies how causation is defined in the model with respect to parameters or properties satisfied by the model.
\item \emph{Inclusion of temporal relationships in the model.} Since the activity of neurons are related over time this condition specifies whether such temporal relationships are incorporated in defining causal relationships among neural activity in the CFC model.
\item \emph{Generalization of the statistical model.} This condition specifies model restrictions, such as linear or non-linear modeling, and informs whether such restrictions can be generalized.
\item \emph{Parametric or non-parametric model.} This specifies whether the model is parametric (i.e. consisting of a parametric equation and needing estimation of parameters) or non-parametric (i.e. free of a parametric equation)~\citep{casella2021statistical}. Non-parametric models have the advantage of not requiring assumptions on specific dynamical equations for the neural dynamics.
\item \emph{Estimation-based vs. hypothesis test-based inference of CFC from recorded data.} Approaches for inferring CFC either support estimation of the CFC from the data or test of significance of hypothetical CFC models based on data. This condition specifies which category among these does the CFC model belong to.
\item \emph{Inclusion of cycles and self-loops in the model.} Neural activity often consist of feedback loops and cycles~\citep{byrne2014molecules}. This condition specifies whether such cycles and self-loops are represented in the CFC model.
\item \emph{Incorporation of intervention queries in a counterfactual manner.} This condition specifies whether interventional queries are answered directly by the CFC model from observational data without performing the experimental intervention.
\item \emph{Ability to recover relationships between neurons when ground-truth dynamic equation ofneural activity are given.} It is often desirable that causal relationships between neural activity in ground truth dynamical equations are accurately represented by the inferred CFC~\citep{schmidt2016multivariate,reid2019advancing}. This condition sheds light into the performance of the CFC approach to recover such ground truth relationships from dynamical equations.
\end{enumerate}
We proceed to delineate causation from statistical associations by surveying existing approaches and describe relation of each approach to above statistical properties~\citep{pearl2000models, pearl2009causal,pearl2009causality}.
\subsection{Granger Causality}
\textit{Granger causality} (GC) is a statistical methodology for determining whether one time series is useful in predicting another \citep{granger1969investigating, basu2015network}. Denoting $X_v(t)$ to be the state of neuron $v$ at time $t$, 
the main idea of GC is that, $X_j$ “Granger-causes” $X_i$ if past of $X_j$ contains information that helps predict the future of $X_i$ better than using the information in the past of $X_i$ or past of other conditioning variables $X_k$~\citep{friston2013analysing}. More formally, typical approach considers a linear Gaussian vector auto-regressive (VAR) linear model between the variables, in this case, states of neurons $X_i$~\citep{lutkepohl2005new},
\[
X_i(t) = \sum_{j=1}^N\sum_{k=1}^K A_{ji}(k)X_j(t - k) + \epsilon_{i}(t)
\]
where $K$ is the maximum number of lags (model order)  and $A_{ji}(K)$ are real-valued linear regression coefficients, and $\epsilon_i (t) \sim N(0,\sigma^2 I)$. Neuron $j$ is said to \emph{Granger-cause} neuron
$i$ if at least one of the coefficients $A_{ji}(k)\neq 0$ for $k = 1,\ldots, K$. In practice, $A_{ji}(k)$ are estimated by minimizing the squared prediction error or by maximizing the likelihood or sparsity-inducing penalized likelihood~\citep{pollonini2010functional,basu2015network}. Granger Causality has been applied towards inference of CFC in the linear Gaussian VAR model setting \citep{guo2020granger, schmidt2016multivariate,pollonini2010functional}. Extensions of GC to categorical random variables, non-linear auto-regressive models and non-parametric models exist \citep{marinazzo2008kernel,dhamala2008analyzing,tank2017granger,tank2018neural}. 

GC was first introduced within econometrics and later has been used to find directed functional connectivity in electrophysiological studies \citep{granger1969investigating,geweke1984measures}, in EEG or MEG datasets, either at source or sensor level \citep{bernasconi1999directionality,ding2000short,brovelli2004beta,barrett2012granger}. The slow dynamics and regional variability of the haemodynamic response to underlying neuronal activity in fMRI were shown to be able to confound the temporal precedence assumptions of GC \citep{david2008identifying,roebroeck2005mapping,wen2012causal,bressler2008top}. 

While Granger Causality provides a powerful tool for understanding which neural time series have a key role in predicting the future of other neural time series \citep{stokes2017study,dahlhaus2003causality,guo2018survey}, studies express concern since prediction is not a formal setting to answer causal questions related to the consequence of interventions and counterfactuals \citep{eichler2013causal,friston2009causal, grassmann2020new}. Furthermore, in practice, GC uses a model assumption between the variables, e.g. a linear Gaussian VAR model, and results could differ when this assumption does not hold \citep{lutkepohl2005new}. Notwithstanding the limitations, GC has been a well-known method in the neural time series scenarios, and applications \citep{guo2020granger,qiao2017functional}. GC is equivalent to Transfer Entropy for Gaussian variables, while the latter is a non-linear method in its general formulation \citep{barnett2009granger}. Transfer Entropy has been explored as a tool to explore connectomics at different scales \citep{ursino2020transfer}, and also applied to the retina circuit \citep{wibral2013measuring}.

\subsection{Dynamic Causal Model}
The Dynamic Causal Model (DCM) is an approach for modeling brain activity and causal interaction between brain regions. DCM was first introduced for fMRI time series and treats the brain as a deterministic non-linear dynamical system network in which neural activity is considered to be unobserved and propagates in an input-state-output system \citep{friston2003dynamic}. The system is attached to a forward model that maps the neural activity to observed Blood Oxygenation Level Dependent (BOLD) fMRI signal \citep{stephan2007comparing,friston2000nonlinear,stephan2007dynamic}. The neural and observational models are specified by a particular parametric form that depends on the data and imaging modality and they together form a fully generative model \citep{friston2013analysing}. DCM outputs evidence for different neural and/or observational models and posterior parameter estimates optimized by variational Bayesian techniques \citep{penny2012comparing}. The coupling parameters between hidden states for different brain regions, in the differential equations that specify the model, constitute the CFC. 
The DCM model incorporates interactions due to the experimental equipment and changes due to external perturbations \citep{marreiros2010dynamic}.

\begin{figure}[htbp]
    \centering
    \includegraphics[width=0.5\textwidth]{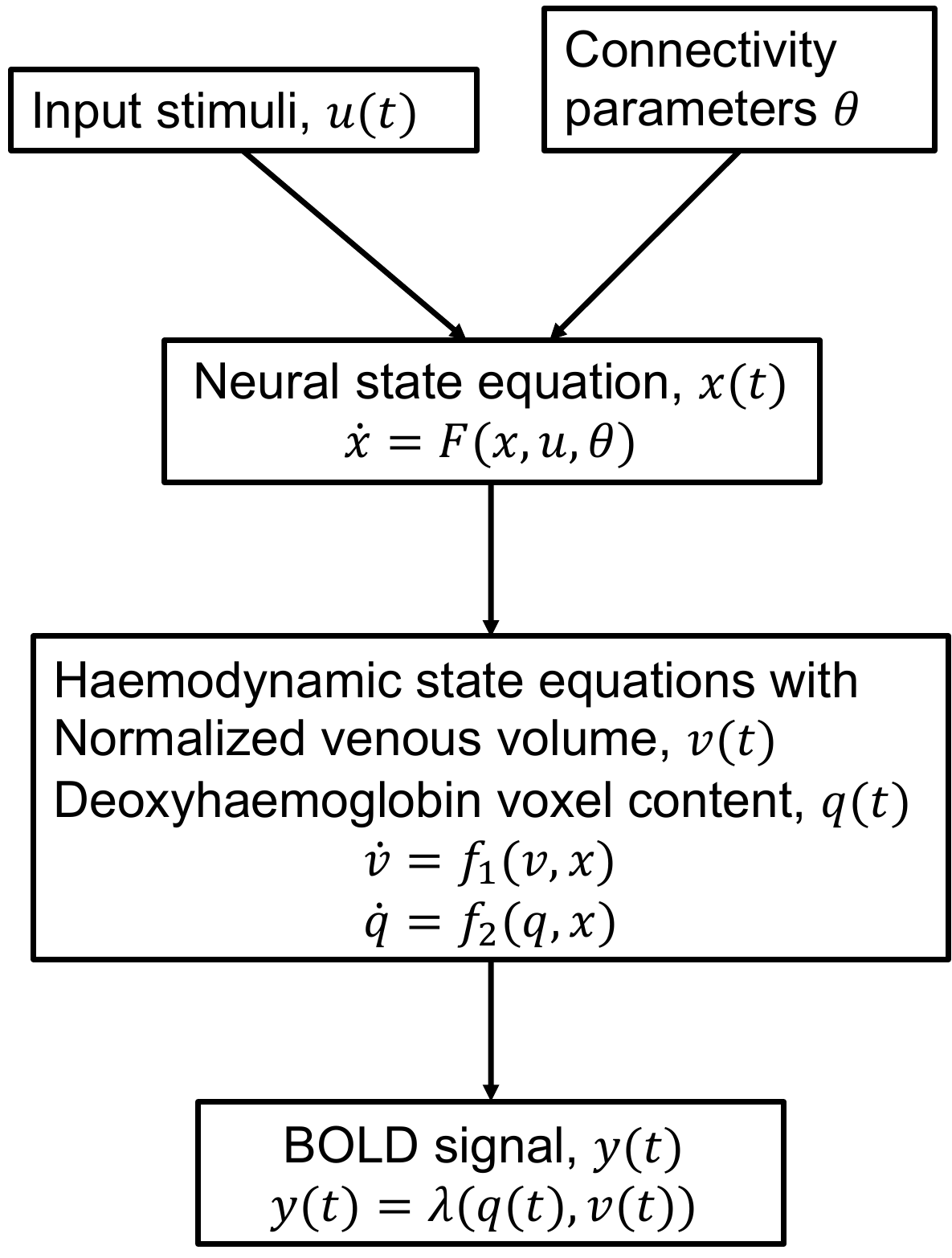}
    \caption{\textbf{Schematic of the haemodynamic model used by DCM for fMRI.} Input stimuli $u(t)$ lead to neural state $x(t)$ subject to connectivity parameters $\theta$ modeled by the neural state equation. Neuronal activity leads to increase in blood flow that changes in venous volume $v(t)$ and deoxyhaemoglobin content $q(t)$ modeled by the haemodynamic state equations. These haemodynamic states are fed to a forward non-linear model $\lambda$ which results in the Blood Oxygenation Level Dependent (BOLD) fMRI response \citep{stephan2007comparing}.}
    \label{fig:dcm}
\end{figure}
DCM compares the evidence for different hypothesized models, and thereby is a tool for testing hypothesis for model and experimental design. DCMs are not restricted to linear systems and include intricate models with large number of parameters for neural dynamics and experimental perturbations. The model is typically constructed to be biologically plausible. Constraints are exercised to the model through priors. The priors are useful to specify which connections are believed to be more likely, e.g., a Bayesian likelihood with the prior distribution being maximized to obtain the parameter estimates. 
The approach relies on precise modeling and aims to specify a biologically plausible detailed mechanistic model of the neuronal dynamics and imaging equipment perturbation \citep{stephan2010ten}. While DCM was originally formulated to be deterministic, recent advances can include stochastic fluctuations in the neural activity as well \citep{li2011generalised,stephan2008nonlinear}. The DCM framework has also been extended beyond fMRI and established in the magneto/encephalography domain \citep{kiebel2008dynamic}, and in local field potentials \citep{moran2013neural}.

\subsection{Directed Probabilistic Graphical Models}\label{sec:dgm} \textit{Directed Probabilistic Graphical Models} (DPGMs) provide a probabilistic foundation to causality in a manner that answers causal queries through a generalized model without requiring specific modeling of the neural dynamics. An important aspect to take into account in inference of CFC is stochasticity. Neural signals are stochastic due to noise and intrinsic nature of neuron firing \citep{stein2005neuronal,manwani1999signal}. The variability and noise in neural dynamics is well known to challenge the determination of neural phenomenon, e.g. the detection of onset of epileptic seizures \citep{vidyaratne2017real, biswas2014peak}. So, when we say ``spike in activity of neuron A is a cause of the spike in activity of neuron B'', the cause makes the effect more likely and not certain due to other factors like incoming inhibitory projections from other neurons and extracellular ion concentration \citep{kroener2009dopamine,soybacs2019real}. 
Moreover, additional arbitrary functional relationships between each cause and effect could exist such that these introduce arbitrary disturbances following some undisclosed probability distribution. For example, it is widely known that diet and stress in humans change the levels of neurotransmitters in the brain \citep{fernstrom1977effects, mora2012stress}. Thereby, the strength of causal relationships between neurons can be perturbed by daily variability in diet and/or stress.   
Also, uncertainties in effect can occur from unobserved causes which is especially true in the context of neural signals as extraneous variables such as diet and stress are often not observed or due to recording from only a fraction of the units in the brain \citep{krishnaswamy2017sparsity}. With these considerations, it is elaborated that values of exogenous variables do not imply values for the remaining variables in a deterministic manner. This motivates the need to consider a probabilistic foundation to causation and causal graphs provided by DPGM. We outline three conditions of DPGM that connect probabilities with causal graphs: The Directed Markov Property, the Causal Minimality Condition, and the Faithfulness Condition.



\begin{figure}
    \centering
    \includegraphics[width=\textwidth]{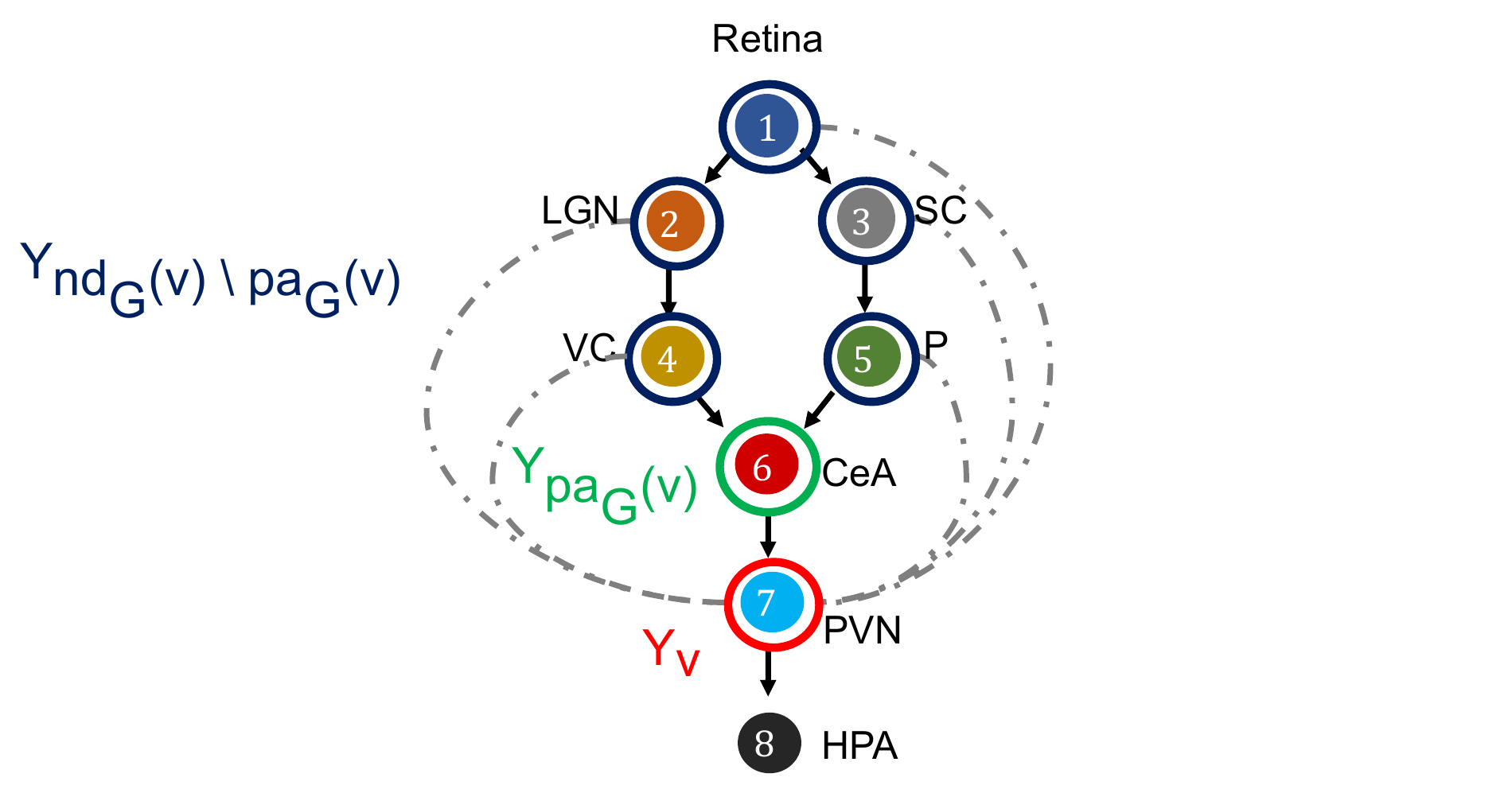}
    \caption{\textbf{Directed Markov Property in the context of fear stimulus.} The DAG in the example of Figure \ref{fig:causal-example} is annotated to illustrate the Directed Markov Property (Eq. \ref{eq:dgmarkov}). PVN is selected as node $v$, its random variable is hence $Y_v$ (red). The parents of v denoted as $pa_G(v)$, and corresponding random variables are $\bm{Y}_{pa_G (v)}$ (green). The non-descendants of $v$ before parents, denoted as $nd_G(v)\setminus pa_G (v)$, and corresponding random variables $\bm{Y}_{nd_G(v)\setminus pa_G (v)}$ (blue). Directed Markov Property holds with the true causal edges (black), as for them parents and children are functionally related (Eq. \ref{eq:causaleff-structural}). Causal Minimality Condition ensures that potential edges (blue dotted) between $nd_G(v)\setminus pa_G(v)$ nodes and $v$ are absent from the DAG.}
    \label{fig:dmp}
\end{figure}

Let $G=(V,E)$ be a DAG over neuron labels $V=(v_1,\ldots,v_N)$ with directed edges $E$ (e.g., Figure \ref{fig:causal-example}). DPGM typically considers the graph to be a DAG because of coherent probability semantics with a DAG and challenges with directed cycles, while there could be more general extensions \citep{lauritzen2001causal,maathuis2018handbook}. Nodes $v$ and $u \in V$ are said to be \emph{adjacent} if $v\rightarrow u \in E$ or $u\rightarrow v \in E$. A \emph{path} is a sequence of distinct nodes in which successive nodes are adjacent. If $\pi = (v_0,\ldots, v_k)$ is a path then $v_0$ and $v_k$ are the end-points of the path. 
If every edge of $\pi$ is of the form $v_{i-1}\rightarrow v_{i}$ then $v_0$ is an \emph{ancestor} of $v_k$ and $v_k$ is a \emph{descendant} of $v_0$. We use the convention that $v$ is an ancestor and descendant of itself. The set of \emph{non-descendants} of $v$, denoted $nd_G(v)$, contains nodes $u\in V$ that are not descendants of $v$. The set of \emph{parents} of $v\in V$ is denoted as $pa_G(v) =\{u\in V: u\rightarrow v\in E\}$. We mark the set $nd_G (v)\setminus pa_G (v)$ as the set that contains all nodes which are older ancestors of $v$ before its parents (Figure \ref{fig:dmp}). Let $Y_v$ denote a scalar-valued random variable corresponding to $v\in V$, e.g., the neural recording at time $t$: $Y_v=X_v(t)$, average of recordings over time $Y_v=\bar{X}(v)$, and for a set of neurons $A\subset V$, $\bm{Y}_A$ denotes the random vector $(Y_v, v\in A)$. With these notations, we outline the three conditions of DPGM.
\paragraph{1) Directed Markov Property}
$(Y_v,v\in V)$ is said to satisfy the \emph{Directed Markov Property} with respect to the DAG $G$ if and only if,
\begin{equation}\label{eq:dgmarkov}
Y_v\ind \bm{Y}_{nd_G (v)\setminus pa_G (v)}|\bm{Y}_{pa_G (v)}
\end{equation}
for every $v\in V$. The Directed Markov Property translates the edges in the DAG into conditional independencies, such that each node $Y_v$ and its older ancestors $\bm{Y}_{nd_G (v)\setminus pa_G (v)}$ are conditionally independent given its parents $\bm{Y}_{pa_G}(v)$. In other words, the influence of each node's ancestors beyond parents reaches to the node exclusively via its parents. In this way, the Directed Markov Property connects probabilistic conditional independencies with relationships of causal influence between nodes of a directed graph. For example, under the Directed Markov Property, in Figure \ref{fig:dmp}, the assertion that the activity of VC and P are conditionally independent of the activity of PVN, given the activity of CeA at time $t$ corresponds to the causal relationship that the influence of the activity of VC and P on the activity of PVN is mediated by the activity of CeA, represented in the DAG as CeA is a parent node of PVN, and VC and P are non-descendant nodes beyond parents of CeA.

The Directed Markov Property for DPGM (DMP) is different from the Markov Property for undirected PGM (MP) in that DMP relates conditional independencies between random variables in a directed graph to causal relationships in the directed graph, whereas MP relates conditional independencies between random variables in an undirected graph to edges of association in the undirected graph. Yet, both incorporate multi-nodal interactions in the graphs beyond a pairwise manner. Furthermore, we had seen in Theorem \ref{thm:hcliff} that MP yields a factorization of the probability density of the random variables comprising the PGM. The DMP also yields a factorization of the joint probability density for DPGM in an adapted manner as follows \citep{verma1988causal}. 
\begin{theorem}[Factorization Theorem for DPGM]
For $(Y_v:v\in V)$ real random variables with density $f$ with respect to a product measure, it satisfies the Directed Markov Property (Eq. \ref{eq:dgmarkov}) with respect to the DAG $G$ if and only if their distribution factorizes according to the $G$, which means, 
\begin{equation}
f(y) = \prod_{v\in V} f(y_v \vert y_{pa_G(v)}), y\in \mathbb{R}^V
\end{equation}
where $f$ is the density of $Y_v$, and $f(y_v \vert y_{pa_G(v)})$ are conditional probability densities.
\end{theorem}

The Directed Markov Property can be equivalently represented with functional relationships between parent and child instead of conditional independencies, which is described in the following theorem \citep{bollen1989structural}. 

\begin{theorem}[Functional Equivalence of DPGM]\label{eq:fnequiv}
If $Y_v$ satisfies
\begin{equation}\label{eq:causaleff-structural}Y_v = g_v(Y_{pa_{\tilde{G}}(v)}, \epsilon_v), v\in V\end{equation}
where $\epsilon_v$ are independent random variables and $g_v$ are measurable functions for $v\in V$ and $\tilde{G}$ is a DAG with vertices $V$, then $Y_v, v\in V$ satisfies the Directed Markov Property with respect to $\tilde{G}$. Conversely, if $Y_v,v\in V$ satisfies the Directed Markov Property with respect to a DAG $\tilde{G}$, then there are
independent random variables $\epsilon_v$ and measurable functions $g_v$ for which Eq. (\ref{eq:causaleff-structural}) holds.
\end{theorem}
Since the functional equations admit a natural causal interpretation, so do DPGMs satisfying the Directed Markov Property \citep{drton2017structure}. The variables in $Y_{pa_{\tilde{G}}}(v)$ are direct causes of $Y_v$, meaning that changes in $Y_{pa_{\tilde{G}}}(v)$ lead to changes in distribution $Y_v$, but not necessarily the other way around. Furthermore, when $Y_v, v\in V$ satisfies the Directed Markov Property, then Eq. (\ref{eq:causaleff-structural}) holds for some choice of functions $\{g_v\}$ and error distributions $\{\epsilon_v\}$, which implies causal relationships among $Y_v, v\in V$. 

DPGMs can predict the consequence of a counterfactual intervention on the random variables \citep{pearl2009causality}. Using Theorem \ref{eq:fnequiv} we show in the following that we only need to remove the edges pointing to the intervened random variables in the DPGM to incorporate the impact of a counterfactual intervention. More precisely, if before the intervention, $G$ is DPGM and $Y_v, v\in V$ satisfies the DMP with respect to $G$, an intervention to the random variables will modify Eq. (\ref{eq:causaleff-structural}) by modifying it for only those variables that are impacted by the intervention. For example, let us consider the intervention forcing $Y_{v_0}$ to take the value 0 or 1 regardless of value at other nodes. This intervention will change Eq. (\ref{eq:causaleff-structural}) by excluding the equations in which $Y_{v_0}$ is a function of other nodes. This corresponds to replacing $pa_{G}(v_0)$ by an empty set in equation Eq. (\ref{eq:causaleff-structural}), and in other words, removing the edges pointing to node $v_0$ in $G$. That is, after the intervention, Eq. (\ref{eq:causaleff-structural}) holds with a different graph $G'$ that is obtained by removing the edges incident upon the intervened nodes in $G$. Equivalently, by Theorem  \ref{eq:fnequiv}, after the intervention $Y_v,v\in V$ satisfies DMP with respect to $G'$, and thus $G'$ is the DPGM after the intervention.

\paragraph{2) Causal Minimality Condition} Let $(Y_v, v\in V)$ satisfy the DMP with respect to the DAG $G$. $G$ satisfies the \emph{Causal Minimality Condition} if and only if for every proper subgraph $H$ of $G$ with vertex set $V$, $(Y_v,v\in V)$ does not satisfy the DMP with respect to $H$. In other words, if adding any edge on to $G$ can also satisfy the DMP, we do not add such an edge, and consider the minimal graph $G$ with respect to which DMP is satisfied to be the DPGM of $Y_v, v\in V$.  In principle, since the complete set of causes is unknown, there can be multiple graphs that would fit a given distribution of random variables for DMP, each connecting the observed variables through different causal relationships. Among all those possible graphs satisfying DMP, the causal minimality condition considers the simplest one and ensures a unique causal DPGM.

\paragraph{3) Faithfulness Condition}
The Directed Markov Property with respect to a DAG $G$ prescribes a set of conditional independence relations on the random variables comprising the graph. However, in general, a probability distribution $P$ of random variables in DAG $G$ that has the independence relations given by the DMP may also include other independence relations. If that does not occur such that all the conditional independence relations by the probability distribution $P$ are encompassed by $G$, then we denote $P$ and $G$ as \emph{faithful} to one another.

\subsubsection{Inference of DPGM}\label{subsec:PCalg}
\begin{figure}[t!]
    \centering
    \includegraphics[width=\textwidth]{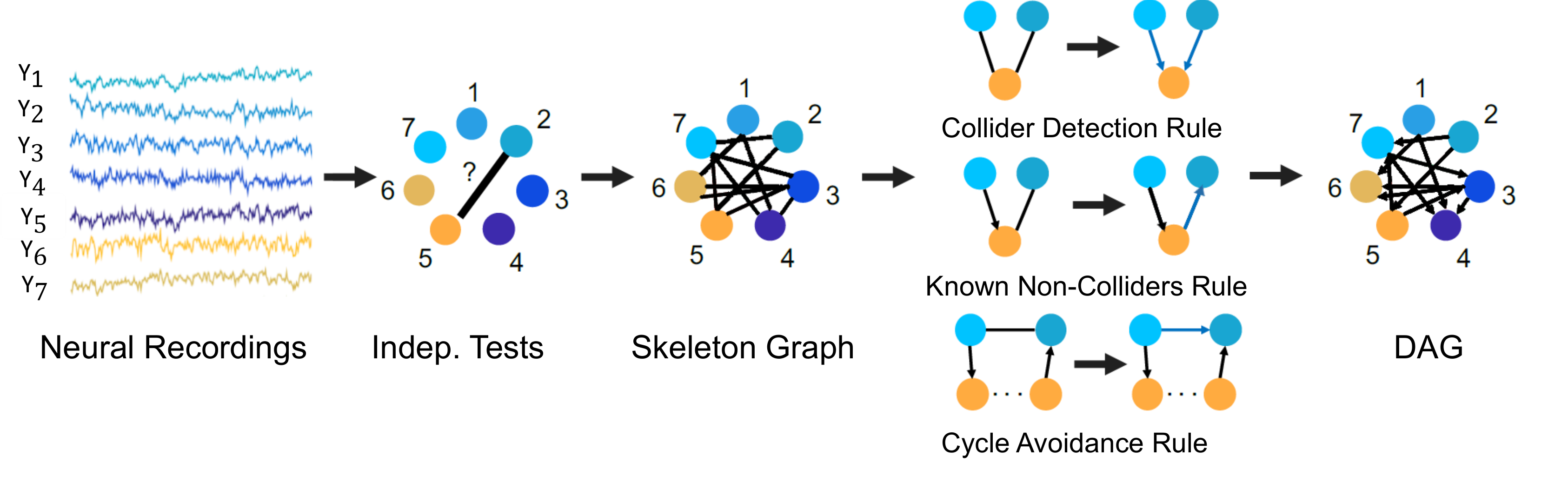}
    \caption{\textbf{The PC algorithm.} Steps of the PC algorithm to infer the DPGM from observed data are summarized by five diagrams (left to right). Data for variables $Y_1-Y_7$ is visualized in the context of neural recordings. Graph with nodes $1-7$ corresponding to variables $Y_1-Y_7$ has no edges. Then, an edge introduced between $Y_i$ and $Y_j$ if they are independent or conditionally independent given any other variable(s) determined by statistical tests, which results in the undirected Skeleton Graph. Using rules such as the Collider Detection rule, Known Non-Colliders rule and Cycle Avoidance rule, the skeleton graph is converted to a DAG.
    }
    \label{fig:fig2}
\end{figure}
Several methods have been developed for inferring DPGM with the Directed Markov Property, Causal Minimality and Faithfulness conditions for stationary observed data. These include the PC algorithm (constraint-based, acyclic graph, no latent confounders, no selection bias) \citep{spirtes2000causation}, FCI algorithm (constraint-based, acyclic graph, latent confounders, selection bias) \citep{spirtes1999algorithm}, GES (score-based equivalence class search) \citep{chickering2002optimal} and GIES (score-based equivalence class search from data with multiple interventions) \citep{hauser2012characterization}.

For example, we describe here the PC algorithm which is a popular statistical tool to infer causal connections between stationary random variables under independent and identically distributed sampling. It is a constraint-based method in which a consistent statistical test for conditional independence is used to select the connectivity graph among the random variables of interest. For Gaussian random variables and linear relationships, a standard choice for such a conditional independence test is constructed using the Fisher's Z-transform \citep{kalisch2007estimating}. For non-Gaussian variables and non-linear relationships, kernel and distance based tests of conditional independence are used (for e.g., the Kernel PC algorithm \citep{tillman2009nonlinear}). The algorithm first represents the observed variables by nodes of a graph and starts with an empty set of edges and decides whether to put an undirected edge between each pair of nodes, for e.g. node 2 and 5 in Figure \ref{fig:fig2}. In order to determine whether to have an edge between the pair of nodes, it performs consistent statistical tests for independence of the random variables for the pair of nodes or conditional independence given the random variables for other node(s). If any of the tests finds evidence for independence or conditional independence, an edge is drawn between the pair of nodes and otherwise no edge is drawn between the pair of nodes. The process is followed for each pair of nodes to result in an undirected graph, called the skeleton graph. Using rules such as the Collider Detection rule, Known Non-Colliders rule and Cycle Avoidance rule, the undirected edges are directed to convert the skeleton graph into a DAG. Figure \ref{fig:fig2} provides a schematic of the algorithm with the context of neural recordings. The PC algorithm is shown to be consistent for finding the true causal graph in the absence of latent confounders \citep{spirtes2000causation}.

\subsection{Comparitive Study of Approaches to Causal Functional Connectivity}
\begin{figure}[t!]
    \centering
    \includegraphics[width=1\textwidth]{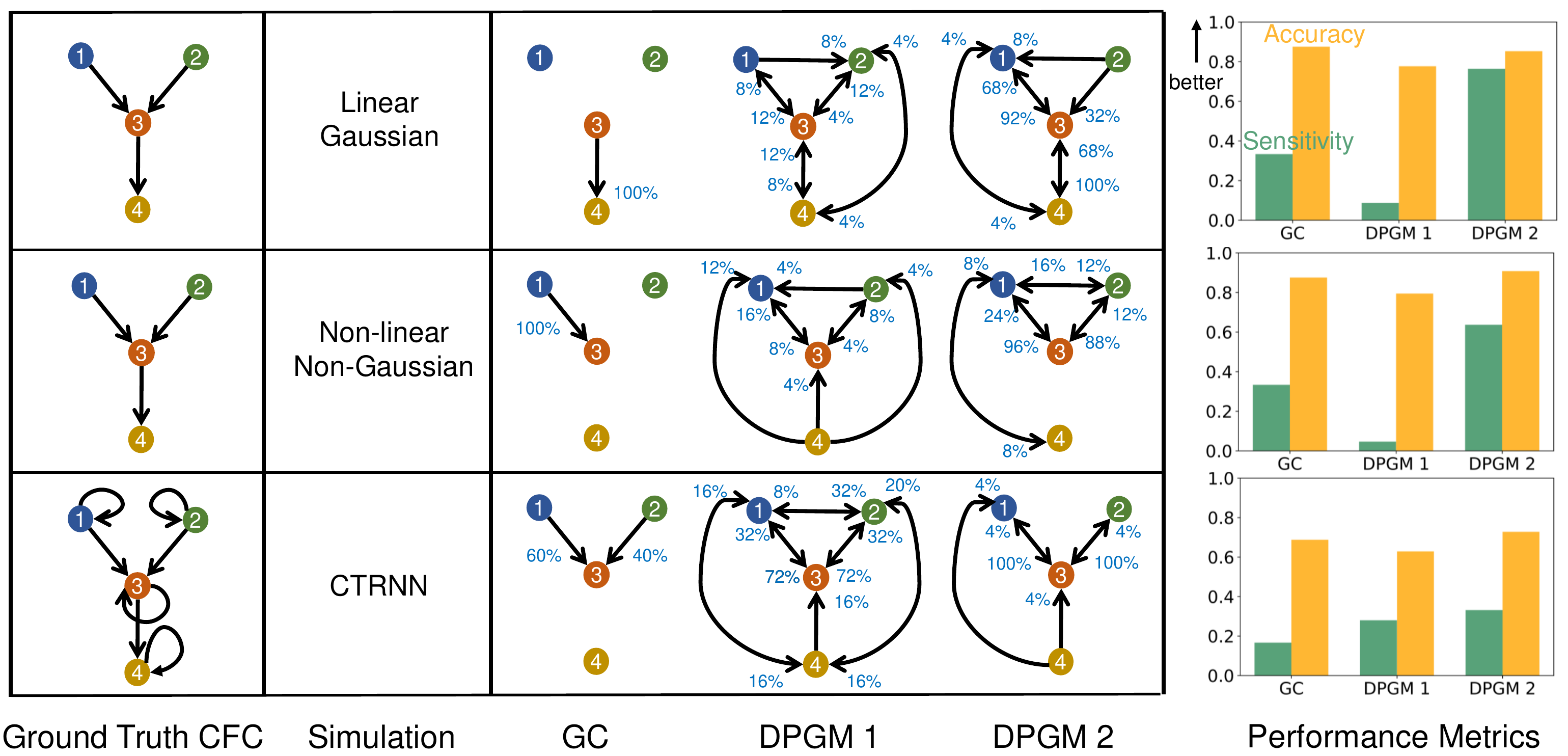}
    \caption{\textbf{Comparative study of CFC inference.} CFC inference of GC, DPGM 1 and DPGM 2 methods is compared on three examples of motifs and simulation paradigms; from top to bottom: Linear Gaussian, Non-linear Non-Gaussian, CTRNN. Table: 4 neurons motifs that define the Ground Truth CFC (left) are depicted side by side  with inferred CFC over several simulation instances according to the three different methods (right). An edge $v\rightarrow w$ in each inferred CFC corresponds to an edge detected in any of the inference instances. The percentage (blue) next to each edge indicates the number of times out of all instances that the edge was detected. Right: For each motif and simulation paradigm, Sensitivity (green) and Accuracy (orange) of each method is shown. 
    }
    \label{fig:simul_eval}
\end{figure}

We compare the performance of exemplary approaches of CFC inference discussed above to recover relationships in ground truth dynamical equations by generating synthetic data from three simulation paradigms and estimate their CFC using the methods of GC and DPGM (see Figure \ref{fig:simul_eval}). The simulation paradigms correspond to specific model assumptions to assess the impact of model assumptions on the performance of the approaches.
\begin{enumerate}[wide=0pt]
\item \underline{Linear Gaussian Time Series (Figure \ref{fig:simul_eval} top-row)}. Let $N(0,1)$ denote a standard Normal random variable. We define $X_v(t)$ as a linear Gaussian time series for $v=1,\ldots,4$ whose true CFC has the edges $1\rightarrow 3,2\rightarrow 3, 3\rightarrow 4$. Let $X_v(0)=N(0,1)$ for $v=1,\ldots,4$ and for $t=1,2,\ldots,10000$,
    \begin{align*}
    &X_1(t)=1+N(0,1),~&&X_2(t)=-1+N(0,1),\\
    &X_3(t+1)=2X_1(t)+X_2(t)+N(0,1),~&&X_4(t+1)=2X_3(t)+N(0,1)
    \end{align*}
    \item \underline{Non-linear Non-Gaussian Time Series (Figure \ref{fig:simul_eval} middle-row)}. Let $U(0,1)$ denote a \emph{Uniformly} distributed random variable on the interval $(0,1)$. We define $X_v(t)$ as a non-linear non-Gaussian time series for $v=1,\ldots,4$ whose true CFC has the edges $1\rightarrow 3, 2\rightarrow 3, 3\rightarrow 4$. Let $X_v(0)=U(0,1)$ for $v=1,\ldots,4$ and for $t=1,2,\ldots,10000$,
    \begin{align*}
    &X_1(t)=U(0,1),~&&X_2(t)=U(0,1),\\
    &X_3(t+1)=4\sin(X_1(t))+3\cos(X_2(t))+U(0,1),~&&X_4(t+1)=2\sin(X_3(t))+U(0,1)
    \end{align*}
    \item \underline{Continuous Time Recurrent Neural Network (CTRNN) (Figure \ref{fig:simul_eval} bottom-row)}. We simulate neural dynamics by Continuous Time Recurrent Neural Networks, Eq. (\ref{ctrnn}). $u_j(t)$ is the instantaneous firing rate at time $t$ for a post-synaptic neuron $j$, $w_{ij}$ is the linear coefficient to pre-synaptic neuron $i$'s input on the post-synaptic neuron $j$, $I_j(t)$ is the input current on neuron $j$ at time $t$, $\tau_j$ is the time constant of the post-synaptic neuron $j$, with $i,j$ being indices for neurons with $m$ being the total number of neurons. Such a model is typically used to simulate neurons as firing rate units.
    \begin{equation}\label{ctrnn}
    \tau_j \frac{du_j(t)}{dt}=-u_j (t) + \sum_{i=1}^m w_{ij} \sigma (u_i (t)) + I_j (t), j=1,\ldots, m
    \end{equation}
    We consider a motif consisting of $4$ neurons with $w_{13}=w_{23}=w_{34}=10$ and $w_{ij}=0$ otherwise. We also note that in Eq. \ref{ctrnn}, activity of each neuron $u_j(t)$ depends on its own past. Therefore, the true CFC has the edges $1\rightarrow 3,2\rightarrow 3,3\rightarrow 4, 1\rightarrow 1, 2\rightarrow 2, 3\rightarrow 3, 4\rightarrow 4$. The time constant $\tau_i$ is set to 10 msecs for each neuron $i$. We consider $I_i(t)$ to be distributed as independent Gaussian process with the mean of 1 and the standard deviation of 1. The signals are sampled at a time gap of $e \approx 2.72$ msecs for a total duration of $10000$ msecs. 
\end{enumerate}

For these network motifs we compare the methods GC, DPGM 1 and DPGM 2. We compute the GC graph using the \emph{Nitime} Python library which fits an MVAR model followed by computation of the Granger Causality by the \emph{GrangerAnalyzer}~\citep{rokem2009nitime}. 

We compute DPGM using the PC algorithm which requires several samples of a scalar-valued random variable $Y_v$ (measured activity) for neurons $v\in V$. We consider two of such $Y_v$ possibilities  

\begin{itemize}[leftmargin=0pt]
    \item[] DPGM 1: Neural recordings at time $t$: $Y_v=X_v(t), v\in V$. Different $t$ gives different samples of $Y_v$. 
   \item[] DPGM 2: Windowed Average of recordings over a duration of $50$ msec: $Y_v=\bar{X}_v, v\in V$, and averaging over different windows of $50$ msec with a gap of $50$ msec in between consecutive windows gives different samples of $Y_v$.
\end{itemize}


We quantified the performance of the algorithms by inference of CFC for 25 different simulations and summarize the performance by two metrics, Accuracy (A) and Sensitivity(S). Let True Positive (TP) be the number of correctly identified edges, True Negative (TN) be the number of missing edges that were correctly identified, False Positive (FP) be the number of incorrectly identified edges and False Negative (FN) be the number of  missing edges incorrectly identified across simulations. We define Accuracy as  \[A=\frac{\text{TP+TN}}{\text{TP+TN+FP+FN}},\] 
which measures the ratio of the count of correctly identified edges or missing edges to the count of all possible edges across simulations. In the motifs and simulation paradigms we consider, there are $4$ neurons and $16$ possible edges (including self-loops) per simulation resulting with total of $400$ possible edges across $25$ simulations. We also define the Sensitivity as
\[S=\frac{\text{TP}}{\text{TP} + \text{FN}}\] 
the ratio of the count of true edges that were correctly identified to the total count of the true edges across simulations. In this comparative study,  Sensitivity is more relevant than Accuracy for the detection of the true edges since it focuses on the detection of the true edges. Indeed, in the extreme case of having the estimated CFC to be an empty set of edges across simulations, the linear Gaussian paradigm will still have $70\%$ neuron pairs correctly identified to be not connected by an edge, thereby resulting in $A=70\%$. Whereas, there will be $0\%$ of true edges detected correctly resulting in $S=0\%$, which reflects the undesirability of the empty graph estimate. We report both Accuracy and Sensitivity for a comprehensive summary of performance. We also report the percentage of the simulations that has each estimated edge present. Higher percentage indicates higher confidence in the detection of that edge. Figure \ref{fig:simul_eval} compares the results for GC, DPGM 1 and DPGM 2 methods in inference of the true CFC. 
\begin{itemize}[leftmargin=0pt]
    \item[] In \textit{Linear Gaussian scenario (top row)}, GC generates a sparse set of edges in which it correctly detects a single edge $3\rightarrow 4$ among the three edges of the true CFC. DPGM 1 generates a large set of edges (9 out of 16 possible) with several of them being spurious. Indeed, each edge is present in less than $25\%$ of the simulations. DPGM 2 generates the same number of edges as DPGM 1, however it has less spurious edges indicated by higher percentages for the expected edges in the Ground Truth CFC ($1\rightarrow 3, 3\rightarrow 4$ with $92\%$ and $100\%$). Overall, all methods result in $A>80\%$ while sensitivity for GC, DPGM 1 and DPGM 2 varied significantly $S=33.3\%, 8.7\%, 76.3\%$ respectively. We thereby conclude that among the three methods, GC is the most accurate but since it did not detect two out of three edges, it is not as sensitive as DPGM 2.
    \item[] In \textit{Non-linear Non-Gaussian scenario (middle row)}, as previously, GC consistently detects a sparse set of edges (single edge $1\rightarrow 3$ with $100\%$) which is one of the three true edges. DPGM 1 obtains a large set of edges with several of them are spurious edges and all edges appear in less than $25\%$ of the trials. For DPGM 2, the number of spurious edges obtained is more than GC and less than DPGM 1. DPGM 2 obtains correctly two out of the three true edges $1\rightarrow 3$ and $2\rightarrow 3$ in most of the trials ($96\%$ and $88\%$ respectively). In summary, GC, DPGM 1 and DPGM 2 resulted in an accuracy of $A=87.5\%, 79.4\%,90.8\%$ respectively and sensitivity of $S=33.3\%, 4.7\%, 63.7\%$ respectively. For this scenario, DPGM 2 has the highest accuracy and sensitivity among the methods. 
    \item[] In \textit{CTRNN scenario (bottom row)}, which includes self-loops, GC obtains two of the three true non-self edges $1\rightarrow 3, 2\rightarrow 3$  for $60\%, 40\%$ of the trials. DPGM 1 detects spurious edges, but also infers the true edges $1\rightarrow 3, 2\rightarrow 3$ for $72\%$ of the trials. DPGM 2 obtains less number of spurious edges compared to DPGM 1 and obtains all of the non-self true edges $1\rightarrow 2$ and $2\rightarrow 3$ for $100\%$ of the trials. In summary, all methods result in a lower accuracy of $A \approx 70\%$ compared to other scenarios since do not include self-loops and sensitivity is $S=16.7\%, 28\%, 33.2\%$ for GC, DPGM 1 and DPGM 2, respectively, which is considerably lower than other scenarios. Among all methods, DPGM 2 obtains the highest accuracy of followed by GC and lastly DPGM 1. DPGM 2 had the highest sensitivity compared to the other methods.
\end{itemize}
 
\section{Discussion}
\begin{table}[t!]
\caption{\textbf{Comparative summary of different approaches for causal modeling in functional connectomics.}} 
\label{causalsummary}
\begin{tabular}{p{3.1cm} p{3.1cm} p{3.1cm} p{3.1cm} p{3.1cm} p{3.1cm}}\toprule
     & \textbf{GC} & \textbf{DCM} & \textbf{DPGM} \\\midrule
    Format of Causality & Non-zero parameters in VAR model & Coupling parameters in biological model & Directed Markov Graph\\\midrule
    
    Inclusion of temporal relationships & \textcolor{ForestGreen}{\textbf{Yes}} & \textcolor{ForestGreen}{\textbf{Yes}} & \textcolor{Maroon}{\textbf{No}}, formulation for stationary variables\\\midrule
    Generalizable statistical model & \textcolor{ForestGreen}{\textbf{Yes}} & \textcolor{Maroon}{\textbf{No}} & \textcolor{ForestGreen}{\textbf{Yes}}\\\midrule
    Non-parametric Model & \textcolor{Maroon}{\textbf{No}}, Linear Gaussian VAR model assumed, there are non-linear extensions. & \textcolor{Maroon}{\textbf{No}}, biologically mechanistic non-linear model. & \textcolor{ForestGreen}{\textbf{Yes}}, equivalent to an arbitrary functional relationship between nodes.\\\midrule
    Supports CFC estimation & \textcolor{ForestGreen}{\textbf{Yes}} & \textcolor{Maroon}{\textbf{No}}, suitable for comparing model hypotheses & \textcolor{ForestGreen}{\textbf{Yes}}\\\midrule
    Occurrence of cycles (including self-loops) in the model & \textcolor{ForestGreen}{\textbf{Yes}} (neuron $i\rightarrow i$ when $A_{ii}(k)\neq 0$ for some $k$) & \textcolor{ForestGreen}{\textbf{Yes}} ($i\rightarrow i$ when $\theta_{ii}\neq 0$) & \textcolor{Maroon}{\textbf{No}}, it is a DAG\\\midrule
    Incorporation of interventional and counterfactual queries & \textcolor{Maroon}{\textbf{No}} & \textcolor{Maroon}{\textbf{No}} & \textcolor{ForestGreen}{\textbf{Yes}} but for stationary variables.\\
    \bottomrule
\end{tabular}
\end{table}
In this paper, we establish a statistical guide to neural connectomics with regards to mapping neural interaction anatomy, function and causality with the means of graphical models. We first describe possibilities of mapping neural anatomical connections with a graph, i.e., anatomical connectome, and demonstrate the difference between such a mapping and a graph that captures functional interactions, i.e., associative functional connectome (AFC). Recognizing that the ultimate goal of functional connectomics is to infer causal interactions between neurons, we define the graphical tools and properties needed to distill AFC into a directional graph, i.e., causal functional connectome (CFC), which represents flows of cause and effect in the interaction between neurons. We then compare exemplary common approaches having the ultimate goal of finding "causation", such as Granger Causality (GC), Dynamic Causal Model (DCM) and Directed Probabilistic Graphical Model (DPGM), in the context of functional connectomics. In particular, we introduce the developments in statistical theory of DPGM to the subject of CFC inference, and define the Directed Markov Property that guarantees consideration of cause and effect in graph settings. We show that this property is key in the definition of probabilistic graphical models that could constitute neural CFC. We then describe the PC algorithm, a common statistical approach for inference of such graphs. Based on these notions and the outcomes of the Directed Markov Property we formulate criteria based on which CFC models can be compared. 
We conclude by performing a holistic comparison, in Table \ref{causalsummary}, of several common approaches that do not obey the Directed Markov Property, such as Granger Causality (GC) and Dynamic Causal Model (DCM), with variants of the PC algorithm (DPGM), comparing them with respect to the criteria that we have outlined. We demonstrate the applicability and the challenges for inference of CFC from measured neural activity for each of the approaches on simulated motifs in Figure~\ref{fig:simul_eval}. 

In this work, our aim is to formulate statistical properties and criteria related to causality of functional connectomics, rather than propose a new approach for causal functional connectome modeling. Such formulation is expected to identify existing gaps in causal modeling and guide extensions of causal functional connectome models ideally satisfying all criteria that we have outlined. Indeed, capturing as many causal criteria is fundamental to any approach from statistical and application points of view. For example, one such property of importance is the ability to uncover directed relationships in ground truth dynamical equations \citep{schmidt2016multivariate,reid2019advancing}, which we include in our comparative study. We have compared the approaches to find CFC in simulations from linear gaussian, non-linear non-gaussian, and CTRNN to demonstrate how specific model assumptions in ground truth dynamical equations are impacting the utilization of the approaches in recovery of relationships between the neurons. For the methods that we have tested, our simulated comparison shows that GC output typically results in a sparse graph with inferred edges that indeed represent causal connections, but we find that it also misses multiple edges that represent causal connections (high accuracy; low sensitivity). For DPGM, we find that the output depends on the choice of measured activity. DPGM 1, which uses full neural time series, results in comparatively low accuracy and low sensitivity since does not capture correlations across time. DPGM 2 uses neural activity averaged over time and results in comparatively high accuracy as well as sensitivity for detection of most of the true edges, but not all of them. These results are not surprising, since the PC algorithm guarantees causality for independent samples per time point thus guarantees the Directed Markov Property only for a single time point, however here we aim to infer a single CFC for the whole recording over time (as in DPGM 1)~\citep{pearl2009causality}. In such a case, the Directed Markov Property and causality are not guaranteed. Averaging over time and separating by time gaps to reduce interdependence between samples leads to improved performance of DPGM 2, but this does not necessarily reflect the time dependent causes and effects between and within neural time series, thus again does not guarantee causality.

Our exposition of properties that each approach is based upon and the comparative study show that each of the methods address different aspects of modeling causality of neural interaction and mapping them in form of a graph. In particular, GC aims to obtain the directed functional connectivity from observed neural states, in a way that tells whether a variable's \emph{past} is predictive of another's future, without requiring a detailed model. It consists of a framework based on auto-regression models which is relatively easy to compute. In contrast, DCM enables \emph{specific} models to be compared based on evidence from data and provides insights on causal connectivity between hidden neural states based on those models. DPGM is a \emph{generic} procedure that represents causal functional connectivity from observed neural states in a way that answers causal queries in both interpretable and mathematically precise manner. To summarize these differences, we outline in Table \ref{causalsummary} the strengths and weaknesses of each of the approaches with respect to applicability to various criteria of causality.  

The comparative table demonstrates that with respect to the model that each approach is assuming, GC requires a linear model in its common use, though has recent non-linear extensions. DCM requires a strict well defined mechanistic biological model and thus can only compare different models based on evidence from data. In comparison, DPGM has an advantage that it does not require modeling of the neural dynamics using a parametric equation or assumption of a linear model. Furthermore, the Directed Markov Condition of the DPGM implies the existence of a functional relationship (Eq. \ref{eq:causaleff-structural}) between parent and children connections in the graph, thus doing away with the need for modeling by specific linear or non-linear functions. Noteworthy, while GC and DCM are parametric, DPGM provides a non-parametric inference for causality and is based on independence tests. In regards to guarantee of causality,  GC can provide useful insights into a system's dynamical interactions in different conditions, however its causal interpretation is not guaranteed as it focuses on the predictability of future based on past observations of variables. DCM uses the parameters for coupling between hidden neural states in competing biological models to indicate CFC, however it compares hypothetical models based on evidence from data which relevance to causality is not guaranteed~\citep{friston2003dynamic}. In comparison, DPGM provides a probabilistic foundation for causality which is predictive of the consequence of possible intervention like neuron ablation and counterfactual queries. Inference of CFC is possible with several causal graph inference algorithms such as the PC algorithm. Such properties of causal interaction between entities are what makes DPGM popular in various disciplines, such as genomics and econometrics~\citep{gomez2020functional,  ahelegbey2016econometrics, ebert2012causal, kalisch2010understanding, deng2005structural, haigh2004causality,wang2017potential,sinoquet2014probabilistic, mourad2012probabilistic, wang2005new, liu2018functional, friedman2004inferring}. However such inference will guarantee causality for stationary measurements only and there is no such guarantee when whole neural time series or their related measurements, such as averaged activity, are considered.

In conclusion, DPGM provides a probabilistic and interpretable formulation for CFC. We have established the statistical properties that the inferred DPGM should posses as well as demonstrated its performance in inference of CFC. While DPGM is a powerful causal framework, existing approaches do not reflect the inter-temporal causal dependencies within and between the neural time series. Yet in the neural time series setting, nodes of the connectivity graph are neurons that correspond to an entire time series of neural activity and comprise inter-temporal dependence. Thus the remaining challenge is the adaptation of the DPGM based formulation of CFC to incorporate inter-temporal dependencies in the neural time series. Such an adaptation will further increase the strength of using DPGM to inferring CFC from neural dynamics.

\bibliographystyle{unsrt}

\begin{thebibliography}{100}

\bibitem{sporns2005human}
Olaf Sporns, Giulio Tononi, and Rolf K{\"o}tter.
\newblock The human connectome: a structural description of the human brain.
\newblock {\em PLoS Comput Biol}, 1(4):e42, 2005.

\bibitem{shi2017connectome}
Yonggang Shi and Arthur~W Toga.
\newblock Connectome imaging for mapping human brain pathways.
\newblock {\em Molecular psychiatry}, 22(9):1230--1240, 2017.

\bibitem{sarwar2020towards}
Tabinda Sarwar, Caio Seguin, Kotagiri Ramamohanarao, and Andrew Zalesky.
\newblock Towards deep learning for connectome mapping: A block decomposition
  framework.
\newblock {\em NeuroImage}, 212:116654, 2020.

\bibitem{xu2020connectome}
C~Shan Xu, Michal Januszewski, Zhiyuan Lu, Shin-ya Takemura, Kenneth Hayworth,
  Gary Huang, Kazunori Shinomiya, Jeremy Maitin-Shepard, David Ackerman, and
  Stuart Berg.
\newblock A connectome of the adult drosophila central brain.
\newblock {\em BioRxiv}, 2020.

\bibitem{lee2011specificity}
Wei-Chung~Allen Lee and R~Clay Reid.
\newblock Specificity and randomness: structure--function relationships in
  neural circuits.
\newblock {\em Current opinion in neurobiology}, 21(5):801--807, 2011.

\bibitem{kopell2014beyond}
Nancy~J Kopell, Howard~J Gritton, Miles~A Whittington, and Mark~A Kramer.
\newblock Beyond the connectome: the dynome.
\newblock {\em Neuron}, 83(6):1319--1328, 2014.

\bibitem{kim2017neural}
Jimin Kim, William Leahy, and Eli Shlizerman.
\newblock Neural interactome: Interactive simulation of a neuronal system.
\newblock {\em Frontiers in Computational Neuroscience}, 13:8, 2019.

\bibitem{kim2018movingcelegans}
Jimin Kim, Julia~A Santos, Mark~J Alkema, and Eli Shlizerman.
\newblock Whole integration of neural connectomics, dynamics and bio-mechanics
  for identification of behavioral sensorimotor pathways in caenorhabditis
  elegans.
\newblock {\em bioRxiv}, page 724328, 2019.

\bibitem{bargmann2013connectome}
Cornelia~I Bargmann and Eve Marder.
\newblock From the connectome to brain function.
\newblock {\em Nature methods}, 10(6):483, 2013.

\bibitem{reid2012functional}
R~Clay Reid.
\newblock From functional architecture to functional connectomics.
\newblock {\em Neuron}, 75(2):209--217, 2012.

\bibitem{hassabis2017neuroscience}
Demis Hassabis, Dharshan Kumaran, Christopher Summerfield, and Matthew
  Botvinick.
\newblock Neuroscience-inspired artificial intelligence.
\newblock {\em Neuron}, 95(2):245--258, 2017.

\bibitem{finn2015functional}
Emily~S Finn, Xilin Shen, Dustin Scheinost, Monica~D Rosenberg, Jessica Huang,
  Marvin~M Chun, Xenophon Papademetris, and R~Todd Constable.
\newblock Functional connectome fingerprinting: identifying individuals using
  patterns of brain connectivity.
\newblock {\em Nature neuroscience}, 18(11):1664--1671, 2015.

\bibitem{shlizerman2012neural}
Eli Shlizerman, Konrad Schroder, and J~Nathan Kutz.
\newblock Neural activity measures and their dynamics.
\newblock {\em SIAM Journal on Applied Mathematics}, 72(4):1260--1291, 2012.

\bibitem{rogers2007assessing}
Baxter~P Rogers, Victoria~L Morgan, Allen~T Newton, and John~C Gore.
\newblock Assessing functional connectivity in the human brain by fmri.
\newblock {\em Magnetic resonance imaging}, 25(10):1347--1357, 2007.

\bibitem{preti2017dynamic}
Maria~Giulia Preti, Thomas~AW Bolton, and Dimitri Van De~Ville.
\newblock The dynamic functional connectome: State-of-the-art and perspectives.
\newblock {\em Neuroimage}, 160:41--54, 2017.

\bibitem{xu2015dynamic}
Yuting Xu and Martin~A Lindquist.
\newblock Dynamic connectivity detection: an algorithm for determining
  functional connectivity change points in fmri data.
\newblock {\em Frontiers in neuroscience}, 9:285, 2015.

\bibitem{wee2016diagnosis}
Chong-Yaw Wee, Pew-Thian Yap, and Dinggang Shen.
\newblock Diagnosis of autism spectrum disorders using temporally distinct
  resting-state functional connectivity networks.
\newblock {\em CNS neuroscience \& therapeutics}, 22(3):212--219, 2016.

\bibitem{varoquaux2010brain}
Gael Varoquaux, Alexandre Gramfort, Jean-Baptiste Poline, and Bertrand Thirion.
\newblock Brain covariance selection: better individual functional connectivity
  models using population prior.
\newblock In {\em Advances in neural information processing systems}, pages
  2334--2342, 2010.

\bibitem{smith2011network}
Stephen~M Smith, Karla~L Miller, Gholamreza Salimi-Khorshidi, Matthew Webster,
  Christian~F Beckmann, Thomas~E Nichols, Joseph~D Ramsey, and Mark~W Woolrich.
\newblock Network modelling methods for fmri.
\newblock {\em Neuroimage}, 54(2):875--891, 2011.

\bibitem{friedman2008sparse}
Jerome Friedman, Trevor Hastie, and Robert Tibshirani.
\newblock Sparse inverse covariance estimation with the graphical lasso.
\newblock {\em Biostatistics}, 9(3):432--441, 2008.

\bibitem{valdes2011effective}
Pedro~A Valdes-Sosa, Alard Roebroeck, Jean Daunizeau, and Karl Friston.
\newblock Effective connectivity: influence, causality and biophysical
  modeling.
\newblock {\em Neuroimage}, 58(2):339--361, 2011.

\bibitem{ramsey2010six}
Joseph~D Ramsey, Stephen~Jos{\'e} Hanson, Catherine Hanson, Yaroslav~O
  Halchenko, Russell~A Poldrack, and Clark Glymour.
\newblock Six problems for causal inference from fmri.
\newblock {\em neuroimage}, 49(2):1545--1558, 2010.

\bibitem{white1986structure}
John~G White, Eileen Southgate, J~Nichol Thomson, and Sydney Brenner.
\newblock The structure of the nervous system of the nematode caenorhabditis
  elegans.
\newblock {\em Philos Trans R Soc Lond B Biol Sci}, 314(1165):1--340, 1986.

\bibitem{conturo1999tracking}
Thomas~E Conturo, Nicolas~F Lori, Thomas~S Cull, Erbil Akbudak, Abraham~Z
  Snyder, Joshua~S Shimony, Robert~C McKinstry, Harold Burton, and Marcus~E
  Raichle.
\newblock Tracking neuronal fiber pathways in the living human brain.
\newblock {\em Proceedings of the National Academy of Sciences},
  96(18):10422--10427, 1999.

\bibitem{le2003looking}
Denis Le~Bihan.
\newblock Looking into the functional architecture of the brain with diffusion
  mri.
\newblock {\em Nature reviews neuroscience}, 4(6):469--480, 2003.

\bibitem{catani2003occipito}
Marco Catani, Derek~K Jones, Rosario Donato, and Dominic~H Ffytche.
\newblock Occipito-temporal connections in the human brain.
\newblock {\em Brain}, 126(9):2093--2107, 2003.

\bibitem{ragan2012serial}
Timothy Ragan, Lolahon~R Kadiri, Kannan~Umadevi Venkataraju, Karsten Bahlmann,
  Jason Sutin, Julian Taranda, Ignacio Arganda-Carreras, Yongsoo Kim,
  H~Sebastian Seung, and Pavel Osten.
\newblock Serial two-photon tomography for automated ex vivo mouse brain
  imaging.
\newblock {\em Nature methods}, 9(3):255--258, 2012.

\bibitem{morone2019symmetry}
Flaviano Morone and Hern{\'a}n~A Makse.
\newblock Symmetry group factorization reveals the structure-function relation
  in the neural connectome of caenorhabditis elegans.
\newblock {\em Nature communications}, 10(1):1--13, 2019.

\bibitem{kato2015global}
Saul Kato, Harris~S Kaplan, Tina Schr{\"o}del, Susanne Skora, Theodore~H
  Lindsay, Eviatar Yemini, Shawn Lockery, and Manuel Zimmer.
\newblock Global brain dynamics embed the motor command sequence of
  caenorhabditis elegans.
\newblock {\em Cell}, 163(3):656--669, 2015.

\bibitem{zatka2020perceptual}
Peter Zatka-Haas, Nicholas~A Steinmetz, Matteo Carandini, and Kenneth~D Harris.
\newblock A perceptual decision requires sensory but not action coding in mouse
  cortex.
\newblock {\em bioRxiv}, page 501627, 2020.

\bibitem{villette2019ultrafast}
Vincent Villette, Mariya Chavarha, Ivan~K Dimov, Jonathan Bradley, Lagnajeet
  Pradhan, Benjamin Mathieu, Stephen~W Evans, Simon Chamberland, Dongqing Shi,
  Renzhi Yang, et~al.
\newblock Ultrafast two-photon imaging of a high-gain voltage indicator in
  awake behaving mice.
\newblock {\em Cell}, 179(7):1590--1608, 2019.

\bibitem{steinmetz2019distributed}
Nicholas~A Steinmetz, Peter Zatka-Haas, Matteo Carandini, and Kenneth~D Harris.
\newblock Distributed coding of choice, action and engagement across the mouse
  brain.
\newblock {\em Nature}, 576(7786):266--273, 2019.

\bibitem{steinmetz2018challenges}
Nicholas~A Steinmetz, Christof Koch, Kenneth~D Harris, and Matteo Carandini.
\newblock Challenges and opportunities for large-scale electrophysiology with
  neuropixels probes.
\newblock {\em Current opinion in neurobiology}, 50:92--100, 2018.

\bibitem{stocco2019analysis}
Andrea Stocco, Zoe Steine-Hanson, Natalie Koh, John~E Laird, Christian~J
  Lebiere, and Paul Rosenbloom.
\newblock Analysis of the human connectome data supports the notion of a
  “common model of cognition” for human and human-like intelligence.
\newblock {\em BioRxiv}, page 703777, 2019.

\bibitem{van2012human}
David~C Van~Essen, Kamil Ugurbil, Edward Auerbach, Deanna Barch, Timothy~EJ
  Behrens, Richard Bucholz, Acer Chang, Liyong Chen, Maurizio Corbetta,
  Sandra~W Curtiss, et~al.
\newblock The human connectome project: a data acquisition perspective.
\newblock {\em Neuroimage}, 62(4):2222--2231, 2012.

\bibitem{jarrell2012connectome}
Travis~A Jarrell, Yi~Wang, Adam~E Bloniarz, Christopher~A Brittin, Meng Xu,
  J~Nichol Thomson, Donna~G Albertson, David~H Hall, and Scott~W Emmons.
\newblock The connectome of a decision-making neural network.
\newblock {\em science}, 337(6093):437--444, 2012.

\bibitem{varshney2011structural}
Lav~R Varshney, Beth~L Chen, Eric Paniagua, David~H Hall, and Dmitri~B
  Chklovskii.
\newblock Structural properties of the caenorhabditis elegans neuronal network.
\newblock {\em PLoS computational biology}, 7(2):e1001066, 2011.

\bibitem{wainwright2008graphical}
Martin~J Wainwright and Michael~Irwin Jordan.
\newblock {\em Graphical models, exponential families, and variational
  inference}.
\newblock Now Publishers Inc, 2008.

\bibitem{drton2017structure}
Mathias Drton and Marloes~H Maathuis.
\newblock Structure learning in graphical modeling.
\newblock {\em Annual Review of Statistics and Its Application}, 4:365--393,
  2017.

\bibitem{epskamp2018gaussian}
Sacha Epskamp, Lourens~J Waldorp, Ren{\'e} M{\~o}ttus, and Denny Borsboom.
\newblock The gaussian graphical model in cross-sectional and time-series data.
\newblock {\em Multivariate Behavioral Research}, 53(4):453--480, 2018.

\bibitem{dyrba2020gaussian}
Martin Dyrba, Reza Mohammadi, Michel~J Grothe, Thomas Kirste, and Stefan~J
  Teipel.
\newblock Gaussian graphical models reveal inter-modal and inter-regional
  conditional dependencies of brain alterations in alzheimer's disease.
\newblock {\em Frontiers in aging neuroscience}, 12:99, 2020.

\bibitem{mlegaussiangraphical}
T.~P. Speed and H.~T. Kiiveri.
\newblock {Gaussian Markov Distributions over Finite Graphs}.
\newblock {\em The Annals of Statistics}, 14(1):138 -- 150, 1986.

\bibitem{shojaie2010penalized}
Ali Shojaie and George Michailidis.
\newblock Penalized likelihood methods for estimation of sparse
  high-dimensional directed acyclic graphs.
\newblock {\em Biometrika}, 97(3):519--538, 2010.

\bibitem{banerjee2008model}
Onureena Banerjee, Laurent~El Ghaoui, and Alexandre d’Aspremont.
\newblock Model selection through sparse maximum likelihood estimation for
  multivariate gaussian or binary data.
\newblock {\em Journal of Machine learning research}, 9(Mar):485--516, 2008.

\bibitem{rothman2008sparse}
Adam~J Rothman, Peter~J Bickel, Elizaveta Levina, Ji~Zhu, et~al.
\newblock Sparse permutation invariant covariance estimation.
\newblock {\em Electronic Journal of Statistics}, 2:494--515, 2008.

\bibitem{meinshausen2006high}
Nicolai Meinshausen, Peter B{\"u}hlmann, et~al.
\newblock High-dimensional graphs and variable selection with the lasso.
\newblock {\em The annals of statistics}, 34(3):1436--1462, 2006.

\bibitem{friston2011functional}
Karl~J Friston.
\newblock Functional and effective connectivity: a review.
\newblock {\em Brain connectivity}, 1(1):13--36, 2011.

\bibitem{horwitz2003elusive}
Barry Horwitz.
\newblock The elusive concept of brain connectivity.
\newblock {\em Neuroimage}, 19(2):466--470, 2003.

\bibitem{yeo2011organization}
BT~Thomas Yeo, Fenna~M Krienen, Jorge Sepulcre, Mert~R Sabuncu, Danial
  Lashkari, Marisa Hollinshead, Joshua~L Roffman, Jordan~W Smoller, Lilla
  Z{\"o}llei, Jonathan~R Polimeni, et~al.
\newblock The organization of the human cerebral cortex estimated by intrinsic
  functional connectivity.
\newblock {\em Journal of neurophysiology}, 106(3):1125, 2011.

\bibitem{power2011functional}
Jonathan~D Power, Alexander~L Cohen, Steven~M Nelson, Gagan~S Wig, Kelly~Anne
  Barnes, Jessica~A Church, Alecia~C Vogel, Timothy~O Laumann, Fran~M Miezin,
  Bradley~L Schlaggar, et~al.
\newblock Functional network organization of the human brain.
\newblock {\em Neuron}, 72(4):665--678, 2011.

\bibitem{power2013control}
Jonathan~D Power and Steven~E Petersen.
\newblock Control-related systems in the human brain.
\newblock {\em Current opinion in neurobiology}, 23(2):223--228, 2013.

\bibitem{smith2009correspondence}
Stephen~M Smith, Peter~T Fox, Karla~L Miller, David~C Glahn, P~Mickle Fox,
  Clare~E Mackay, Nicola Filippini, Kate~E Watkins, Roberto Toro, Angela~R
  Laird, et~al.
\newblock Correspondence of the brain's functional architecture during
  activation and rest.
\newblock {\em Proceedings of the National Academy of Sciences},
  106(31):13040--13045, 2009.

\bibitem{carr2015ll}
James~A Carr.
\newblock I'll take the low road: the evolutionary underpinnings of visually
  triggered fear.
\newblock {\em Frontiers in neuroscience}, 9:414, 2015.

\bibitem{pessoa2010emotion}
Luiz Pessoa and Ralph Adolphs.
\newblock Emotion processing and the amygdala: from a'low road'to'many roads'
  of evaluating biological significance.
\newblock {\em Nature reviews neuroscience}, 11(11):773--782, 2010.

\bibitem{bertini2013blind}
Caterina Bertini, Roberto Cecere, and Elisabetta L{\`a}davas.
\newblock I am blind, but i “see” fear.
\newblock {\em Cortex}, 49(4):985--993, 2013.

\bibitem{morris1999subcortical}
John~S Morris, Arne {\"O}hman, and Raymond~J Dolan.
\newblock A subcortical pathway to the right amygdala mediating “unseen”
  fear.
\newblock {\em Proceedings of the National Academy of Sciences},
  96(4):1680--1685, 1999.

\bibitem{horn2020opportunities}
Andreas Horn and Michael~D Fox.
\newblock Opportunities of connectomic neuromodulation.
\newblock {\em Neuroimage}, 221:117180, 2020.

\bibitem{pearl2009causality}
Judea Pearl.
\newblock {\em Causality}.
\newblock Cambridge university press, 2009.

\bibitem{casella2021statistical}
George Casella and Roger~L Berger.
\newblock {\em Statistical inference}.
\newblock Cengage Learning, 2021.

\bibitem{byrne2014molecules}
John~H Byrne, Ruth Heidelberger, and M~Neal Waxham.
\newblock {\em From molecules to networks: an introduction to cellular and
  molecular neuroscience}.
\newblock Academic Press, 2014.

\bibitem{schmidt2016multivariate}
Christoph Schmidt, Britta Pester, Nicole Schmid-Hertel, Herbert Witte, Axel
  Wism{\"u}ller, and Lutz Leistritz.
\newblock A multivariate granger causality concept towards full brain
  functional connectivity.
\newblock {\em PloS one}, 11(4):e0153105, 2016.

\bibitem{reid2019advancing}
Andrew~T Reid, Drew~B Headley, Ravi~D Mill, Ruben Sanchez-Romero, Lucina~Q
  Uddin, Daniele Marinazzo, Daniel~J Lurie, Pedro~A Vald{\'e}s-Sosa,
  Stephen~Jos{\'e} Hanson, Bharat~B Biswal, et~al.
\newblock Advancing functional connectivity research from association to
  causation.
\newblock {\em Nature neuroscience}, 1(10), 2019.

\bibitem{pearl2000models}
Judea Pearl et~al.
\newblock Models, reasoning and inference.
\newblock {\em Cambridge, UK: CambridgeUniversityPress}, 2000.

\bibitem{pearl2009causal}
Judea Pearl et~al.
\newblock Causal inference in statistics: An overview.
\newblock {\em Statistics surveys}, 3:96--146, 2009.

\bibitem{granger1969investigating}
Clive~WJ Granger.
\newblock Investigating causal relations by econometric models and
  cross-spectral methods.
\newblock {\em Econometrica: journal of the Econometric Society}, pages
  424--438, 1969.

\bibitem{basu2015network}
Sumanta Basu, Ali Shojaie, and George Michailidis.
\newblock Network granger causality with inherent grouping structure.
\newblock {\em The Journal of Machine Learning Research}, 16(1):417--453, 2015.

\bibitem{friston2013analysing}
Karl Friston, Rosalyn Moran, and Anil~K Seth.
\newblock Analysing connectivity with granger causality and dynamic causal
  modelling.
\newblock {\em Current opinion in neurobiology}, 23(2):172--178, 2013.

\bibitem{lutkepohl2005new}
Helmut L{\"u}tkepohl.
\newblock {\em New introduction to multiple time series analysis}.
\newblock Springer Science \& Business Media, 2005.

\bibitem{pollonini2010functional}
Luca Pollonini, Udit Patidar, Ning Situ, Roozbeh Rezaie, Andrew~C Papanicolaou,
  and George Zouridakis.
\newblock Functional connectivity networks in the autistic and healthy brain
  assessed using granger causality.
\newblock In {\em 2010 Annual International Conference of the IEEE Engineering
  in Medicine and Biology}, pages 1730--1733. IEEE, 2010.

\bibitem{guo2020granger}
Xinling Guo, Qiaosheng Zhang, Amrita Singh, Jing Wang, and Zhe~Sage Chen.
\newblock Granger causality analysis of rat cortical functional connectivity in
  pain.
\newblock {\em Journal of Neural Engineering}, 17(1):016050, 2020.

\bibitem{marinazzo2008kernel}
Daniele Marinazzo, Mario Pellicoro, and Sebastiano Stramaglia.
\newblock Kernel method for nonlinear granger causality.
\newblock {\em Physical review letters}, 100(14):144103, 2008.

\bibitem{dhamala2008analyzing}
Mukeshwar Dhamala, Govindan Rangarajan, and Mingzhou Ding.
\newblock Analyzing information flow in brain networks with nonparametric
  granger causality.
\newblock {\em Neuroimage}, 41(2):354--362, 2008.

\bibitem{tank2017granger}
Alex Tank, Emily~B Fox, and Ali Shojaie.
\newblock Granger causality networks for categorical time series.
\newblock {\em arXiv preprint arXiv:1706.02781}, 2017.

\bibitem{tank2018neural}
Alex Tank, Ian Covert, Nicholas Foti, Ali Shojaie, and Emily Fox.
\newblock Neural granger causality for nonlinear time series.
\newblock {\em arXiv preprint arXiv:1802.05842}, 2018.

\bibitem{geweke1984measures}
John~F Geweke.
\newblock Measures of conditional linear dependence and feedback between time
  series.
\newblock {\em Journal of the American Statistical Association},
  79(388):907--915, 1984.

\bibitem{bernasconi1999directionality}
Corrado Bernasconi and Peter Ko{\`E}nig.
\newblock On the directionality of cortical interactions studied by structural
  analysis of electrophysiological recordings.
\newblock {\em Biological cybernetics}, 81(3):199--210, 1999.

\bibitem{ding2000short}
Mingzhou Ding, Steven~L Bressler, Weiming Yang, and Hualou Liang.
\newblock Short-window spectral analysis of cortical event-related potentials
  by adaptive multivariate autoregressive modeling: data preprocessing, model
  validation, and variability assessment.
\newblock {\em Biological cybernetics}, 83(1):35--45, 2000.

\bibitem{brovelli2004beta}
Andrea Brovelli, Mingzhou Ding, Anders Ledberg, Yonghong Chen, Richard
  Nakamura, and Steven~L Bressler.
\newblock Beta oscillations in a large-scale sensorimotor cortical network:
  directional influences revealed by granger causality.
\newblock {\em Proceedings of the National Academy of Sciences},
  101(26):9849--9854, 2004.

\bibitem{barrett2012granger}
Adam~B Barrett, Michael Murphy, Marie-Aur{\'e}lie Bruno, Quentin Noirhomme,
  M{\'e}lanie Boly, Steven Laureys, and Anil~K Seth.
\newblock Granger causality analysis of steady-state electroencephalographic
  signals during propofol-induced anaesthesia.
\newblock {\em PloS one}, 7(1):e29072, 2012.

\bibitem{david2008identifying}
Olivier David, Isabelle Guillemain, Sandrine Saillet, Sebastien Reyt, Colin
  Deransart, Christoph Segebarth, and Antoine Depaulis.
\newblock Identifying neural drivers with functional mri: an
  electrophysiological validation.
\newblock {\em PLoS Biol}, 6(12):e315, 2008.

\bibitem{roebroeck2005mapping}
Alard Roebroeck, Elia Formisano, and Rainer Goebel.
\newblock Mapping directed influence over the brain using granger causality and
  fmri.
\newblock {\em Neuroimage}, 25(1):230--242, 2005.

\bibitem{wen2012causal}
Xiaotong Wen, Li~Yao, Yijun Liu, and Mingzhou Ding.
\newblock Causal interactions in attention networks predict behavioral
  performance.
\newblock {\em Journal of Neuroscience}, 32(4):1284--1292, 2012.

\bibitem{bressler2008top}
Steven~L Bressler, Wei Tang, Chad~M Sylvester, Gordon~L Shulman, and Maurizio
  Corbetta.
\newblock Top-down control of human visual cortex by frontal and parietal
  cortex in anticipatory visual spatial attention.
\newblock {\em Journal of Neuroscience}, 28(40):10056--10061, 2008.

\bibitem{stokes2017study}
Patrick~A Stokes and Patrick~L Purdon.
\newblock A study of problems encountered in granger causality analysis from a
  neuroscience perspective.
\newblock {\em Proceedings of the national academy of sciences},
  114(34):E7063--E7072, 2017.

\bibitem{dahlhaus2003causality}
Rainer Dahlhaus and Michael Eichler.
\newblock Causality and graphical models in time series analysis.
\newblock {\em Oxford Statistical Science Series}, pages 115--137, 2003.

\bibitem{guo2018survey}
Ruocheng Guo, Lu~Cheng, Jundong Li, P~Richard Hahn, and Huan Liu.
\newblock A survey of learning causality with data: Problems and methods.
\newblock {\em arXiv preprint arXiv:1809.09337}, 2018.

\bibitem{eichler2013causal}
Michael Eichler.
\newblock Causal inference with multiple time series: principles and problems.
\newblock {\em Philosophical Transactions of the Royal Society A: Mathematical,
  Physical and Engineering Sciences}, 371(1997):20110613, 2013.

\bibitem{friston2009causal}
Karl Friston.
\newblock Causal modelling and brain connectivity in functional magnetic
  resonance imaging.
\newblock {\em PLoS biology}, 7(2), 2009.

\bibitem{grassmann2020new}
Greta Grassmann.
\newblock New considerations on the validity of the wiener-granger causality
  test.
\newblock {\em Heliyon}, 6(10):e05208, 2020.

\bibitem{qiao2017functional}
Jianping Qiao, Zhishun Wang, Guihu Zhao, Yuankai Huo, Carl~L Herder, Chamonix~O
  Sikora, and Bradley~S Peterson.
\newblock Functional neural circuits that underlie developmental stuttering.
\newblock {\em PLoS One}, 12(7):e0179255, 2017.

\bibitem{barnett2009granger}
Lionel Barnett, Adam~B Barrett, and Anil~K Seth.
\newblock Granger causality and transfer entropy are equivalent for gaussian
  variables.
\newblock {\em Physical review letters}, 103(23):238701, 2009.

\bibitem{ursino2020transfer}
Mauro Ursino, Giulia Ricci, and Elisa Magosso.
\newblock Transfer entropy as a measure of brain connectivity: a critical
  analysis with the help of neural mass models.
\newblock {\em Frontiers in computational neuroscience}, 14:45, 2020.

\bibitem{wibral2013measuring}
Michael Wibral, Nicolae Pampu, Viola Priesemann, Felix Siebenh{\"u}hner, Hannes
  Seiwert, Michael Lindner, Joseph~T Lizier, and Raul Vicente.
\newblock Measuring information-transfer delays.
\newblock {\em PloS one}, 8(2):e55809, 2013.

\bibitem{friston2003dynamic}
Karl~J Friston, Lee Harrison, and Will Penny.
\newblock Dynamic causal modelling.
\newblock {\em Neuroimage}, 19(4):1273--1302, 2003.

\bibitem{stephan2007comparing}
Klaas~Enno Stephan, Nikolaus Weiskopf, Peter~M Drysdale, Peter~A Robinson, and
  Karl~J Friston.
\newblock Comparing hemodynamic models with dcm.
\newblock {\em Neuroimage}, 38(3):387--401, 2007.

\bibitem{friston2000nonlinear}
Karl~J Friston, Andrea Mechelli, Robert Turner, and Cathy~J Price.
\newblock Nonlinear responses in fmri: the balloon model, volterra kernels, and
  other hemodynamics.
\newblock {\em NeuroImage}, 12(4):466--477, 2000.

\bibitem{stephan2007dynamic}
Klaas~E Stephan, Lee~M Harrison, Stefan~J Kiebel, Olivier David, Will~D Penny,
  and Karl~J Friston.
\newblock Dynamic causal models of neural system dynamics: current state and
  future extensions.
\newblock {\em Journal of biosciences}, 32(1):129--144, 2007.

\bibitem{penny2012comparing}
William~D Penny.
\newblock Comparing dynamic causal models using aic, bic and free energy.
\newblock {\em Neuroimage}, 59(1):319--330, 2012.

\bibitem{marreiros2010dynamic}
Andr{\'e}~C Marreiros, Stefan~J Kiebel, and Karl~J Friston.
\newblock A dynamic causal model study of neuronal population dynamics.
\newblock {\em Neuroimage}, 51(1):91--101, 2010.

\bibitem{stephan2010ten}
Klaas~Enno Stephan, Will~D Penny, Rosalyn~J Moran, Hanneke~EM den Ouden, Jean
  Daunizeau, and Karl~J Friston.
\newblock Ten simple rules for dynamic causal modeling.
\newblock {\em Neuroimage}, 49(4):3099--3109, 2010.

\bibitem{li2011generalised}
Baojuan Li, Jean Daunizeau, Klaas~E Stephan, Will Penny, Dewen Hu, and Karl
  Friston.
\newblock Generalised filtering and stochastic dcm for fmri.
\newblock {\em neuroimage}, 58(2):442--457, 2011.

\bibitem{stephan2008nonlinear}
Klaas~Enno Stephan, Lars Kasper, Lee~M Harrison, Jean Daunizeau, Hanneke~EM den
  Ouden, Michael Breakspear, and Karl~J Friston.
\newblock Nonlinear dynamic causal models for fmri.
\newblock {\em Neuroimage}, 42(2):649--662, 2008.

\bibitem{kiebel2008dynamic}
Stefan~J Kiebel, Marta~I Garrido, Rosalyn~J Moran, and Karl~J Friston.
\newblock Dynamic causal modelling for eeg and meg.
\newblock {\em Cognitive neurodynamics}, 2(2):121--136, 2008.

\bibitem{moran2013neural}
Rosalyn~J Moran, Dimitris~A Pinotsis, and Karl~J Friston.
\newblock Neural masses and fields in dynamic causal modeling.
\newblock {\em Frontiers in computational neuroscience}, 7:57, 2013.

\bibitem{stein2005neuronal}
Richard~B Stein, E~Roderich Gossen, and Kelvin~E Jones.
\newblock Neuronal variability: noise or part of the signal?
\newblock {\em Nature Reviews Neuroscience}, 6(5):389--397, 2005.

\bibitem{manwani1999signal}
Amit Manwani and Christof Koch.
\newblock Signal detection in noisy weakly-active dendrites.
\newblock {\em Advances in neural information processing systems}, pages
  132--138, 1999.

\bibitem{vidyaratne2017real}
Lasitha~S Vidyaratne and Khan~M Iftekharuddin.
\newblock Real-time epileptic seizure detection using eeg.
\newblock {\em IEEE Transactions on Neural Systems and Rehabilitation
  Engineering}, 25(11):2146--2156, 2017.

\bibitem{biswas2014peak}
Rahul Biswas, Koulik Khamaru, and Kaushik~K Majumdar.
\newblock A peak synchronization measure for multiple signals.
\newblock {\em IEEE Transactions on Signal Processing}, 62(17):4390--4398,
  2014.

\bibitem{kroener2009dopamine}
Sven Kroener, L~Judson Chandler, Paul~EM Phillips, and Jeremy~K Seamans.
\newblock Dopamine modulates persistent synaptic activity and enhances the
  signal-to-noise ratio in the prefrontal cortex.
\newblock {\em PloS one}, 4(8):e6507, 2009.

\bibitem{soybacs2019real}
Zafer Soyba{\c{s}}, Sefa {\c{S}}im{\c{s}}ek, FM~Bet{\"u}l Erol, U~{\c{C}}iya
  Erdo{\u{g}}an, Esra~N {\c{S}}im{\c{s}}ek, B{\"u}{\c{s}}ra {\c{S}}ahin, Merve
  Mar{\c{c}}al{\i}, Bahattin Aydo{\u{g}}du, {\c{C}}a{\u{g}}lar Elb{\"u}ken, and
  Rohat Melik.
\newblock Real-time in vivo control of neural membrane potential by
  electro-ionic modulation.
\newblock {\em Iscience}, 17:347--358, 2019.

\bibitem{fernstrom1977effects}
John~D Fernstrom.
\newblock Effects of the diet on brain neurotransmitters.
\newblock {\em Metabolism}, 26(2):207--223, 1977.

\bibitem{mora2012stress}
Francisco Mora, Gregorio Segovia, Alberto Del~Arco, Marta de~Blas, and Pedro
  Garrido.
\newblock Stress, neurotransmitters, corticosterone and body--brain
  integration.
\newblock {\em Brain research}, 1476:71--85, 2012.

\bibitem{krishnaswamy2017sparsity}
Pavitra Krishnaswamy, Gabriel Obregon-Henao, Jyrki Ahveninen, Sheraz Khan,
  Behtash Babadi, Juan~Eugenio Iglesias, Matti~S H{\"a}m{\"a}l{\"a}inen, and
  Patrick~L Purdon.
\newblock Sparsity enables estimation of both subcortical and cortical activity
  from meg and eeg.
\newblock {\em Proceedings of the National Academy of Sciences},
  114(48):E10465--E10474, 2017.

\bibitem{lauritzen2001causal}
Steffen~L Lauritzen.
\newblock Causal inference from graphical models.
\newblock {\em Complex stochastic systems}, pages 63--107, 2001.

\bibitem{maathuis2018handbook}
Marloes Maathuis, Mathias Drton, Steffen Lauritzen, and Martin Wainwright.
\newblock {\em Handbook of graphical models}.
\newblock CRC Press, 2018.

\bibitem{verma1988causal}
T~Verma and J~Pearl.
\newblock Causal networks: Semantics and expressiveness, in proceedings.
\newblock In {\em 4th Workshop on Uncertainty in Artificial Intelligence,
  Minneapolis, MN, Mountain View, CA}, page 352, 1988.

\bibitem{bollen1989structural}
Kenneth~A Bollen.
\newblock Structural equations with latent variables wiley.
\newblock {\em New York}, 1989.

\bibitem{spirtes2000causation}
Peter Spirtes, Clark~N Glymour, Richard Scheines, and David Heckerman.
\newblock {\em Causation, prediction, and search}.
\newblock MIT press, 2000.

\bibitem{spirtes1999algorithm}
Peter Spirtes, Christopher Meek, and Thomas Richardson.
\newblock An algorithm for causal inference in the presence of latent variables
  and selection bias.
\newblock {\em Computation, causation, and discovery}, 21:1--252, 1999.

\bibitem{chickering2002optimal}
David~Maxwell Chickering.
\newblock Optimal structure identification with greedy search.
\newblock {\em Journal of machine learning research}, 3(Nov):507--554, 2002.

\bibitem{hauser2012characterization}
Alain Hauser and Peter B{\"u}hlmann.
\newblock Characterization and greedy learning of interventional markov
  equivalence classes of directed acyclic graphs.
\newblock {\em The Journal of Machine Learning Research}, 13(1):2409--2464,
  2012.

\bibitem{kalisch2007estimating}
Markus Kalisch and Peter B{\"u}hlmann.
\newblock Estimating high-dimensional directed acyclic graphs with the
  pc-algorithm.
\newblock {\em Journal of Machine Learning Research}, 8(Mar):613--636, 2007.

\bibitem{tillman2009nonlinear}
Robert~E Tillman, Arthur Gretton, and Peter Spirtes.
\newblock Nonlinear directed acyclic structure learning with weakly additive
  noise models.
\newblock In {\em NIPS}, pages 1847--1855. Citeseer, 2009.

\bibitem{rokem2009nitime}
Ariel Rokem, M~Trumpis, and F~Perez.
\newblock Nitime: time-series analysis for neuroimaging data.
\newblock In {\em Proceedings of the 8th Python in Science Conference}, pages
  68--75, 2009.

\bibitem{gomez2020functional}
Ana Mar{\'\i}a~Estrada G{\'o}mez, Kamran Paynabar, and Massimo Pacella.
\newblock Functional directed graphical models and applications in root-cause
  analysis and diagnosis.
\newblock {\em Journal of Quality Technology}, pages 1--17, 2020.

\bibitem{ahelegbey2016econometrics}
Daniel~Felix Ahelegbey.
\newblock The econometrics of bayesian graphical models: a review with
  financial application.
\newblock {\em Journal of Network Theory in Finance}, 2(2):1--33, 2016.

\bibitem{ebert2012causal}
Imme Ebert-Uphoff and Yi~Deng.
\newblock Causal discovery for climate research using graphical models.
\newblock {\em Journal of Climate}, 25(17):5648--5665, 2012.

\bibitem{kalisch2010understanding}
Markus Kalisch, Bernd~AG Fellinghauer, Eva Grill, Marloes~H Maathuis, Ulrich
  Mansmann, Peter B{\"u}hlmann, and Gerold Stucki.
\newblock Understanding human functioning using graphical models.
\newblock {\em BMC Medical Research Methodology}, 10(1):1--10, 2010.

\bibitem{deng2005structural}
Ke~Deng, Delin Liu, Shan Gao, and Zhi Geng.
\newblock Structural learning of graphical models and its applications to
  traditional chinese medicine.
\newblock In {\em International Conference on Fuzzy Systems and Knowledge
  Discovery}, pages 362--367. Springer, 2005.

\bibitem{haigh2004causality}
Michael~S Haigh and David~A Bessler.
\newblock Causality and price discovery: An application of directed acyclic
  graphs.
\newblock {\em The Journal of Business}, 77(4):1099--1121, 2004.

\bibitem{wang2017potential}
Huange Wang, Fred~A van Eeuwijk, and Johannes Jansen.
\newblock The potential of probabilistic graphical models in linkage map
  construction.
\newblock {\em Theoretical and Applied Genetics}, 130(2):433--444, 2017.

\bibitem{sinoquet2014probabilistic}
Christine Sinoquet.
\newblock {\em Probabilistic graphical models for genetics, genomics, and
  postgenomics}.
\newblock OUP Oxford, 2014.

\bibitem{mourad2012probabilistic}
Rapha{\"e}l Mourad, Christine Sinoquet, and Philippe Leray.
\newblock Probabilistic graphical models for genetic association studies.
\newblock {\em Briefings in bioinformatics}, 13(1):20--33, 2012.

\bibitem{wang2005new}
Junbai Wang, Leo Wang-Kit Cheung, and Jan Delabie.
\newblock New probabilistic graphical models for genetic regulatory networks
  studies.
\newblock {\em Journal of biomedical informatics}, 38(6):443--455, 2005.

\bibitem{liu2018functional}
Hexuan Liu, Jimin Kim, and Eli Shlizerman.
\newblock Functional connectomics from neural dynamics: probabilistic graphical
  models for neuronal network of caenorhabditis elegans.
\newblock {\em Philosophical Transactions of the Royal Society B: Biological
  Sciences}, 373(1758):20170377, 2018.

\bibitem{friedman2004inferring}
Nir Friedman.
\newblock Inferring cellular networks using probabilistic graphical models.
\newblock {\em Science}, 303(5659):799--805, 2004.

\end{thebibliography}

\end{document}